\documentclass[12pt]{article}
\pdfoutput=1
\usepackage{color}
\usepackage{amssymb,amsmath}
\usepackage{graphicx}
\usepackage{braket}
\usepackage{epsfig}
\usepackage{here}
\graphicspath{{./Figures/}}
\newcommand{\ba}{\begin{align}}
\newcommand{\ea}{\end{align}}

 \makeatletter
\def\alt{\mathrel{\mathpalette\gl@align<}}
\def\agt{\mathrel{\mathpalette\gl@align>}}
\def\gl@align#1#2{\lower.6ex\vbox{\baselineskip\z@skip\lineskip\z@
\ialign{$\m@th#1\hfil##\hfil$\crcr#2\crcr\sim\crcr}}} \makeatother

\begin{document}
\begin{flushright}
\end{flushright}
\vspace*{1.0cm}

\begin{center}
\baselineskip 20pt 
{\Large\bf 
$\epsilon'/\epsilon$ Anomaly and Neutron EDM in
\\
$SU(2)_L\times SU(2)_R\times U(1)_{B-L}$ model with Charge Symmetry
}
\vspace{1cm}

{\large 
Naoyuki Haba, \ Hiroyuki Umeeda \ and \ Toshifumi Yamada
} \vspace{.5cm}

{\baselineskip 20pt \it
Graduate School of Science and Engineering, Shimane University, Matsue 690-8504, Japan
}

\vspace{.5cm}

\vspace{1.5cm} {\bf Abstract} \end{center}
The Standard Model prediction for $\epsilon'/\epsilon$
based on recent lattice QCD results exhibits a tension with the experimental data.
We solve this tension through $W_R^+$ gauge boson exchange
in the $SU(2)_L\times SU(2)_R\times U(1)_{B-L}$ model with `charge symmetry',
 whose theoretical motivation is to attribute the chiral structure of the Standard Model to
 the spontaneous breaking of $SU(2)_R\times U(1)_{B-L}$ gauge group and charge symmetry.
We show that $M_{W_R}<58$~TeV is required to account for the $\epsilon'/\epsilon$ anomaly in this model.
Next, we make a prediction for the neutron EDM in the same model and 
study a correlation between $\epsilon'/\epsilon$ and the neutron EDM.
We confirm that the model can solve the $\epsilon'/\epsilon$ anomaly without conflicting the current bound
 on the neutron EDM,
 and further reveal that almost all parameter regions in which the $\epsilon'/\epsilon$ anomaly is explained
 will be covered by future neutron EDM searches,
 which leads us to anticipate the discovery of the neutron EDM.

\thispagestyle{empty}

%\bigskip
\newpage

%\addtocounter{page}{-1}
\setcounter{footnote}{0}
%%%%%%%%%%%%%%%%%%%%%%%%%%
%\baselineskip 36pt
% Main body
%%%%%%%%%%%%%%%%%%%%%%%%%%
\baselineskip 18pt
%%%%%%%%%%%%%%%%%%%%%%%%%%
%

\section{Introduction}

The direct CP violation in $K\to\pi\pi$ decay parametrized by the $\epsilon'$ parameter
is sensitive to physics beyond the Standard Model (SM) due to the suppressed SM contribution.
Recent calculation of the hadronic matrix elements with lattice QCD~\cite{Blum:2011ng, Blum:2015ywa, Bai:2015nea}
enables us to evaluate the $K\to\pi\pi$ decay amplitude without relying on any hadron model.
On the basis of the above calculation, the same collaboration has reported that the SM prediction is separated from the experimental value~\cite{Batley:2002gn, AlaviHarati:2002ye, Abouzaid:2010ny} by 2.1$\sigma$,
and other groups~\cite{Buras:2015yba, Kitahara:2016nld}
have also obtained predictions for $\epsilon'/\epsilon$ that show a discrepancy of $2.9\sigma$ and $2.8\sigma$, respectively.
More importantly, the lattice result corroborates the calculation with dual QCD approach~\cite{Buras:2015xba,Buras:2016fys},
 which has derived a theoretical upper bound on $\epsilon'/\epsilon$ that is violated by the experimental data
 and has thus claimed anomaly in this observable.
(However, Ref.~\cite{Gisbert:2017vvj} presents a different calculation that claims the absence of the anomaly.)
Some authors have tackled this $\epsilon'/\epsilon$ anomaly in new physics scenarios,
such as a general right-handed current \cite{Cirigliano:2016yhc},
the Littlest Higgs model with T-parity \cite{Blanke:2015wba},
supersymmetry \cite{Tanimoto:2016yfy, Kitahara:2016otd, Endo:2016aws},
non-standard interaction with $Z^\prime$ and/or $Z$ 
\cite{Buras:2015jaq, Endo:2016tnu},
vector-like quarks \cite{Bobeth:2016llm}
and $SU(3)_c\times SU(3)_L\times U(1)_X$ gauge group \cite{Buras:2015kwd}.

The $SU(2)_L \times SU(2)_R \times U(1)_{B-L}$ gauge extension of the SM
 is a well-motivated framework for addressing the $\epsilon'/\epsilon$ puzzle, 
 because the flavor mixing matrix for right-handed quarks automatically introduces new CP-violating phases,
 and $W_R^+$ gauge boson exchange contributes to $\Delta F=1$ processes at tree level while it contributes to $\Delta F=2$ processes at loop levels
 so that other experimental constraints, in particular the constraint from Re($\epsilon$), are readily evaded.
Previously, Ref.~\cite{Cirigliano:2016yhc} has shown that a general $SU(2)_L \times SU(2)_R \times U(1)_{B-L}$ model with
 an arbitrary right-handed quark mixing can solve the $\epsilon'/\epsilon$ discrepancy.
However, a major theoretical motivation for the $SU(2)_L \times SU(2)_R \times U(1)_{B-L}$ model
 lies in its capability of explaining the origin of the chiral nature of the SM,
 which is achieved by adding either the left-right parity~\cite{lr} or the `charge symmetry'~\cite{chargesym}
\footnote{
The `charge symmetry' is inspired by $D$-parity~\cite{dparity} in the $SO(10)$ grand unification theory.
However, the model we consider cannot be embedded in the $SO(10)$ theory, since we assume the charge symmetry breaking scale to be below $O(100)$~TeV.
}.
The left-right parity requires invariance of the theory 
 under the Lorentzian parity transformation
 plus the exchange of $SU(2)_L$ and $SU(2)_R$ gauge groups, 
 while the charge symmetry requires invariance
 under the charge conjugation plus the exchange of $SU(2)_L$ and $SU(2)_R$,
 both of which endow the model with a symmetric structure for the left and right-handed fermions
 at high energies.

In this paper, we study $\epsilon'/\epsilon$
in the $SU(2)_L \times SU(2)_R \times U(1)_{B-L}$ model with charge symmetry.
As a consequence of the charge symmetry,
the Yukawa matrices are complex symmetric matrices,
which restricts the quark mixing matrix associated with
$W_R^+$ to be the complex conjugate of the SM Cabibbo-Kobayashi-Maskawa (CKM) matrix 
multiplied by a new CP phase factor for each quark flavor.
Given the above restriction,
one can evaluate $\epsilon'/\epsilon$
only in terms of two new CP phases, the mass of $W_R^+$ and the ratio of two vacuum expectation values (VEVs) of
the bifundamental scalar,
which leads to a specific prediction for the model parameters.

Our analysis on $\epsilon'/\epsilon$
 proceeds as follows.
By integrating out $W_R$, $W_L$ and the top quark,
 we obtain the Wilson coefficients for $\Delta S=1$ operators that contribute to
 $K\to\pi\pi$ decay.
The anomalous dimension matrix
 is divided into the same two $18\times 18$ pieces for 36 operators,
 for which leading order expressions are obtainable
 from Refs.~\cite{Cho:1993zb, Ciuchini}.
The hadronic matrix elements for current-current operators
 are seized from the lattice results \cite{Blum:2015ywa, Bai:2015nea}.
We find that
 among new physics operators,
 $(\bar{s} u)_L(\bar{u}d)_R$ and
 $(\bar{s} u)_R(\bar{u}d)_R$ (each with two ways of color contraction)
 both dominantly contribute to $\epsilon'/\epsilon$.
Their contributions are of the same order because the Wilson coefficients of
 the $(\bar{s}u)_L(\bar{u}d)_R$ operators
 are suppressed by the hierarchy of 
 two bifundamental scalar VEVs $v_1/v_2=\tan\beta$, which is about
 $m_b/m_t$ if there is no fine-tuning
 in accommodating the top and bottom quark Yukawa couplings,
 whereas this suppression does not enter into the Wilson coefficients of
 the $(\bar{s}u)_R(\bar{u}d)_R$ operators.
On the other hand, the lattice computation has confirmed that 
 the hadronic matrix elements for the former operators
 are enhanced compared to the latter.
Thus, these operators possibly equally contribute to $\epsilon'/\epsilon$.
This result is in contrast to the study of Ref.~\cite{Cirigliano:2016yhc}, 
 which has concentrated solely on the $(\bar{s}u)_L(\bar{u}d)_R$ operators.

Once the $\epsilon'/\epsilon$ anomaly is explained 
 in the $SU(2)_L \times SU(2)_R \times U(1)_{B-L}$ model with charge symmetry,
 correlated predictions for other CP violating observables are of interest.
In particular, the neutron electric dipole moment (EDM),
 an observable sensitive to CP violation
 in the presence of CPT invariance,
 receives significant contributions from four-quark operators 
 in $SU(2)_L \times SU(2)_R \times U(1)_{B-L}$ models~\cite{Beall:1981zq,An:2009zh,Xu:2009nt,Dekens:2014jka,Maiezza:2014ala,Cirigliano:2016yhc}
\footnote{
See also Refs.~\cite{Zhang:2007fn,Zhang:2007da}.
},
 allowing us to discuss
 future detectability of the neutron EDM
 in relation to the $\epsilon'/\epsilon$ anomaly.

Our analysis on the neutron EDM starts by integrating out $W_R$, $W_L$ and top quark
 to obtain the Wilson coefficients for CP-violating operators.
The leading order expression for the anomalous dimension matrix
 is found in Refs.\ \cite{Shifman:1976de,Dai:1989yh,Boyd:1990bx, Hisano:2012cc}.
Regarding the hadronic matrix elements of CP-violating operators,
 we reveal that the pion VEV $\langle\pi^0\rangle$ induced by four-quark operators~\cite{deVries:2012ab}
 gives the leading contribution to the neutron EDM, which is enhanced by the quark mass ratio $m_s/(m_u+m_d)$
 in comparison to the rest.
This enhancement is understood as follows:
Since $W_R^+$-$W_L^+$ mixing gives rise to CP-odd and isospin-odd interactions,
 the pion VEV $\langle\pi^0\rangle$, which is isospin-odd, can arise without the factor of $m_d-m_u$,
 and thus can be directly proportional to $1/(m_u+m_d)$.
The pion VEV induces a CP-violating coupling for neutron $n$, $\Sigma^-$ baryon, and kaon $K^+$
 without the factor of $m_d-m_u$ because the $\bar{n} \Sigma^- K^+$ vertex is not isospin-even.
Consequently, the CP-violating coupling for $n,\Sigma^-, K^+$ can appear with the factor of $m_s/(m_u+m_d)$.
This coupling contributes to the neutron EDM at the leading chiral order through charged baryon-meson loops.
Considering the above-mentioned importance of the pion VEV,
 we in this paper investigate meson condensation,
 the resultant CP-violating baryon-meson couplings,
 and their contributions to the neutron EDM through baryon-meson loops,
 using chiral perturbation theory.

This paper is organized as follows:
In Sec.\ \ref{Sec:Model}, we review the $SU(2)_L \times SU(2)_R \times U(1)_{B-L}$ model with charge symmetry,
 with emphasis on new sources of CP violation.
In Sec.\ \ref{Sec:epsilonprime}, we present the Wilson coefficients for $\Delta S=1$ operators in the model,
 their RG evolutions and the hadronic matrix elements for these operators.
The numerical result for $\epsilon'/\epsilon$ is shown at the end of the section.
In Sec. \ref{Sec:EDM}, we give the Wilson coefficients 
 for CP-violating operators contributing to the neutron EDM.
Special care is taken in evaluating meson condensates and their impact on the neutron EDM.
The final result is a prediction for the neutron EDM in light of
the $\epsilon'/\epsilon$ anomaly.
Section \ref{Summary} is devoted to summary and discussions.
\\

\section{$SU(2)_L\times SU(2)_R\times U(1)_{B-L}$ Model with Charge Symmetry}
\label{Sec:Model}

We consider $SU(3)_C \times SU(2)_L \times SU(2)_R \times U(1)_{B-L}$ gauge theory with charge symmetry.
The field content is in Table~\ref{content}.
\begin{table}[h]
\caption{Field content and charge assignments. $i$ labels the three generations.}
\begin{center}
\begin{tabular}{|c|c|c|c|c|c|c|} \hline
Field               & Lorentz $SO(1,3)$ & $SU(3)_C$ & $SU(2)_L$  &  $SU(2)_R$  & $U(1)_{B-L}$ \\ \hline \hline
$Q_L^i$        & ({\bf 2},\,{\bf 1})  &  {\bf 3}        & {\bf 2}          &  {\bf 1}         &1/3                 \\ 
$Q_R^{c\,i}$& ({\bf 2},\,{\bf 1})  & $\bar{\bf 3}$   & {\bf 1}          &  {\bf 2}         &  $-1/3$           \\ 
$L_L^i$     & ({\bf 2},\,{\bf 1})  &{\bf 1}          & {\bf 2}          &  {\bf 1}         &  $-1$               \\
$L_R^{c\,i}$& ({\bf 2},\,{\bf 1}) &{\bf 1}        & {\bf 1}          &  {\bf 2}         &  $1$          \\ \hline
$\Phi$                & {\bf 1}                   &{\bf 1}        &  {\bf 2}         &  {\bf 2}          &  0      \\
$\Delta_L$      & {\bf 1}                     &{\bf 1}       &  {\bf 3}          &  {\bf 1}         & $2$ \\ 
$\Delta_R$      & {\bf 1}                     &{\bf 1}       &  {\bf 1}          &  {\bf 3}         & $-2$ \\ \hline 
\end{tabular}
\end{center}
\label{content}
\end{table}
Hereafter, the fields are expressed in a way that they transform under a $SU(2)_L\times SU(2)_R$ gauge transformation as
\begin{align} % requires amsmath; align* for no eq. number
\Phi&\to e^{i\theta_L^a\tau^a} \Phi e^{-i\theta_R^b\tau^b},
\ \ \ \Delta_L\to e^{i\theta_L^a\tau^a} \Delta_L e^{-i\theta_L^a\tau^a},
\ \ \ \Delta_R\to (e^{i\theta_R^b\tau^b})^* \Delta_R (e^{-i\theta_R^b\tau^b})^*, \nonumber \\
Q_L^i&\to e^{i\theta_L^a\tau^a}Q_L^i, \ \ \ Q_R^{c\,i}\to (e^{i\theta_R^b\tau^b})^*Q_R^{c\,i}, \ \ \
L_L^i\to e^{i\theta_L^a\tau^a}L_L^i, \ \ \ L_R^{c\,i}\to (e^{i\theta_R^b\tau^b})^*L_R^{c\,i},
\end{align}
 with $\theta_L^a$ and $\theta_R^a$ being gauge parameters for $SU(2)_L$ and $SU(2)_R$, respectively.
We demand the theory to be invariant under the following `charge symmetry' transformation:
\begin{align} % requires amsmath; align* for no eq. number
&{\rm charge \ conjugation \ of \ all \ gauge \ fields },\nonumber \\
&{\rm and} \ SU(2)_L\leftrightarrow SU(2)_R, \ \ \ Q_L^i\leftrightarrow Q_R^{c\,i}, \ \ \ L_L^i\leftrightarrow L_R^{c\,i}, \ \ \ \Phi\leftrightarrow \Phi^T, \ \ \ \Delta_L\leftrightarrow\Delta_R.
\label{chargesym}
\end{align}

The part of the Lagrangian describing $SU(2)_L\times SU(2)_R\times U(1)_{B-L}$ and Yukawa interactions of quarks is given by
\begin{align} % requires amsmath; align* for no eq. number
-{\cal L}&\supset Q_L^{i\,\dagger} \, \bar{\sigma}_\mu\left( \frac{1}{2}g_L \sigma^a W_L^{a\,\mu}+\frac{1}{3}g_X X^\mu\right)Q_L^i
+Q_R^{c\,i\,\dagger} \, \bar{\sigma}_\mu\left( -\frac{1}{2}g_R (\sigma^a)^T W_R^{a\,\mu}-\frac{1}{3}g_X X^\mu\right)Q_R^{c\,i}
\nonumber\\
&+(Y_q)_{ij}\,Q_L^{i\,\dagger}\Phi\epsilon_s(Q_R^{c\,j})^*+(\tilde{Y}_q)_{ij}\,Q_L^{i\,\dagger}(\epsilon_g^T\Phi^*\epsilon_g)\epsilon_s(Q_R^{c\,j})^*+{\rm H.c.},
\label{lag}
\end{align}
 where $g_L$, $g_R$ and $g_X$ are the gauge coupling constants for $SU(2)_L$, $SU(2)_R$ and $U(1)_{B-L}$
 gauge groups, respectively, and $Y_q$ and $\tilde{Y}_q$ are the quark Yukawa couplings.
$\epsilon_s$ denotes the antisymmetric tensor for Lorentz spinors and $\epsilon_g$ denotes that for the fundamental representation of $SU(2)_L$ or $SU(2)_R$.
Invariance under the charge symmetry transformation Eq.~(\ref{chargesym}) leads to the following tree-level relations:
\begin{align} % requires amsmath; align* for no eq. number
g_L&=g_R, \ \ \ (Y_q)_{ij}=(Y_q)_{ji}, \ \ \ (\tilde{Y}_q)_{ij}=(\tilde{Y}_q)_{ji}.
\end{align}

The $SU(2)_R$ triplet scalar $\Delta_R$ develops a VEV, $v_R$, to break $SU_R(2)\times U(1)_{B-L} \to U(1)_Y$,
 and the bi-fundamental scalar $\Phi$ takes a VEV configuration,
\begin{align} % requires amsmath; align* for no eq. number
\langle\Phi\rangle&= \frac{1}{\sqrt{2}}\begin{pmatrix} % or pmatrix or bmatrix or Bmatrix or ...
      v\sin\beta & 0 \\
      0 & v\cos\beta e^{i\,\alpha} \\
   \end{pmatrix},
\end{align}
 to break $SU(2)_L\times U(1)_Y \to U(1)_{em}$, where $\alpha$ is the spontaneous CP phase.
The VEV of $\Delta_L$ is hereafter neglected, as it is severely constrained from $\rho$-parameter.
The resultant mass matrices for $W_L^a$, $W_R^a$ and $X$ gauge bosons read
\begin{align} % requires amsmath; align* for no eq. number
-{\cal L}&\supset\begin{pmatrix} % or pmatrix or bmatrix or Bmatrix or ...
      W_L^- & W_R^- \\
   \end{pmatrix}
      \begin{pmatrix} % or pmatrix or bmatrix or Bmatrix or ...
         g_L^2\ v^2/4 & -g_Lg_R\ \sin(2\beta)e^{-i\,\alpha}v^2/4\\
         -g_Lg_R\ \sin(2\beta)e^{i\,\alpha}v^2/4 & g_R^2(v_R^2+v^2/4) \\
      \end{pmatrix}
         \begin{pmatrix} % or pmatrix or bmatrix or Bmatrix or ...
            W_L^+ \\
            W_R^+ \\
         \end{pmatrix}
         \nonumber \\
         &+
         \frac{1}{2}   \begin{pmatrix} % or pmatrix or bmatrix or Bmatrix or ...
              W_L^3 & W_R^3 & X \\
            \end{pmatrix} 
              \begin{pmatrix} % or pmatrix or bmatrix or Bmatrix or ...
                g_L^2\ v^2/4 & -g_Lg_R\ v^2/2 & 0 \\
                -g_Lg_R\ v^2/2 & g_R^2(2v_R^2+v^2/4) & -2g_Lg_X\ v_R^2 \\
                0 & -2g_Lg_X\ v_R^2 & 2g_X^2\ v_R^2
              \end{pmatrix}
                 \begin{pmatrix} % or pmatrix or bmatrix or Bmatrix or ...
                    W_L^3 \\
                    W_R^3 \\
                    X \\
                 \end{pmatrix}.
\end{align}
The mass matrix for the charged gauge bosons is diagonalized as
\begin{align} % requires amsmath; align* for no eq. number
-{\cal L}\supset\begin{pmatrix} % or pmatrix or bmatrix or Bmatrix or ...
      W^- & W'^- \\
   \end{pmatrix}
      \begin{pmatrix} % or pmatrix or bmatrix or Bmatrix or ...
         M_W^2 & 0 \\
         0 & M_{W'}^2 \\
      \end{pmatrix}
         \begin{pmatrix} % or pmatrix or bmatrix or Bmatrix or ...
            W^+ \\
            W'^+ \\
         \end{pmatrix},
\ \ \ \ \ &\begin{pmatrix} % or pmatrix or bmatrix or Bmatrix or ...
      W_L^+ \\
      W_R^+ \\
   \end{pmatrix}=   \begin{pmatrix} % or pmatrix or bmatrix or Bmatrix or ...
         \cos\zeta &-e^{-i\,\alpha}\sin\zeta  \\
         e^{i\,\alpha}\sin\zeta  & \cos\zeta \\
      \end{pmatrix}
         \begin{pmatrix} % or pmatrix or bmatrix or Bmatrix or ...
            W^+ \\
            W'^+ \\
         \end{pmatrix},
\nonumber \\
&\sin(2\zeta)=\frac{2g_Lg_R \sin(2\beta)v^2}{(g_L^2+g_R^2)v^2+4g_R^2v_R^2-8M_W^2}.
\label{wlwr}
\end{align}
For $v_R\gg v$ and $g_L=g_R$, we have an important relation for $\zeta$,
\begin{align} % requires amsmath; align* for no eq. number
\sin\zeta&\simeq\sin(2\beta)\frac{M_W^2}{M_{W'}^2},
\end{align}
 which indicates that when we assume $\tan\beta\simeq m_b/m_t$ so that the top and bottom Yukawa couplings are naturally derived, 
 the $W_L$-$W_R$ mixing angle $\zeta$ is smaller than $M_W^2/M_{W'}^2$ by the factor $2m_b/m_t\sim0.05$.

The quark mass matrices are given by
\footnote{
$U_R^i\equiv \epsilon_s (U_R^{c\,i})^*$, $D_R^i\equiv \epsilon_s (D_R^{c\,i})^*$.
}
\begin{align} % requires amsmath; align* for no eq. number
-{\cal L}&\supset (M_u)_{ij}\,U_L^{i\,\dagger}U_R^j+(M_d)_{ij}\,D_L^{i\,\dagger}D_R^j+{\rm H.c.},
\nonumber \\
M_u&=\frac{v}{\sqrt{2}}\left(\sin\beta Y_q+\cos\beta e^{-i\,\alpha}\tilde{Y}_q\right), \ \ \
M_d=\frac{v}{\sqrt{2}}\left(\cos\beta e^{i\,\alpha}Y_q+\sin\beta\tilde{Y}_q\right),
\end{align}
 which we diagonalize as
 $M_u=V_{uL}^\dagger {\rm diag}(m_u,\,m_c,\,m_t) V_{uR}$ and $M_d=V_{dL}^\dagger {\rm diag}(m_d,\,m_s,\,m_b) V_{dR}$, with $V_{uL},V_{uR},V_{dL},V_{dR}$ being unitary matrices.
However, since $Y_q$ and $\tilde{Y}_q$ are complex symmetric matrices, so are $M_u$ and $M_d$, and
 one can most generally write
\begin{align} % requires amsmath; align* for no eq. number
V_{uR}&=   \begin{pmatrix} % or pmatrix or bmatrix or Bmatrix or ...
      e^{i\,\phi_u} & 0 & 0 \\
      0 & e^{i\,\phi_c} & 0 \\
      0 & 0 & e^{i\,\phi_t} \\
   \end{pmatrix}V_{uL}^*, \ \ \ \ \
   V_{dR}=   \begin{pmatrix} % or pmatrix or bmatrix or Bmatrix or ...
      e^{i\,\psi_d} & 0 & 0 \\
      0 & e^{i\,\psi_s} & 0 \\
      0 & 0 & e^{i\,\psi_b} \\
   \end{pmatrix}V_{dL}^*.
\end{align}
Hence, the SM CKM matrix, $V_L=V_{uL}V_{dL}^\dagger$, and the corresponding flavor mixing matrix for right-handed quarks, $V_R=V_{uR}V_{dR}^\dagger$,
 are related as
\begin{align} % requires amsmath; align* for no eq. number
V_R=V_{uR}V_{dR}^\dagger&=\begin{pmatrix} % or pmatrix or bmatrix or Bmatrix or ...
      e^{i\,\phi_u} & 0 & 0 \\
      0 & e^{i\,\phi_c} & 0 \\
      0 & 0 & e^{i\,\phi_t} \\
   \end{pmatrix}V_{uL}^*V_{dL}^T\begin{pmatrix} % or pmatrix or bmatrix or Bmatrix or ...
      e^{-i\,\psi_d} & 0 & 0 \\
      0 & e^{-i\,\psi_s} & 0 \\
      0 & 0 & e^{-i\,\psi_b} \\
   \end{pmatrix}
\nonumber \\
&=\begin{pmatrix} % or pmatrix or bmatrix or Bmatrix or ...
      e^{i\,\phi_u} & 0 & 0 \\
      0 & e^{i\,\phi_c} & 0 \\
      0 & 0 & e^{i\,\phi_t} \\
   \end{pmatrix}V_L^*\begin{pmatrix} % or pmatrix or bmatrix or Bmatrix or ...
      e^{-i\,\psi_d} & 0 & 0 \\
      0 & e^{-i\,\psi_s} & 0 \\
      0 & 0 & e^{-i\,\psi_b} \\
   \end{pmatrix}.\label{Eq:RCKM}
\end{align}
Eventually, the part of the Lagrangian Eq.~(\ref{lag}) describing flavor-changing $W,W'$ interactions is recast, in the unitary gauge, into the form,
\begin{align} % requires amsmath; align* for no eq. number
-{\cal L}&\supset\frac{g_L}{\sqrt{2}} (V_L)_{ij} \, U_L^{i\,\dagger} \, W_L^{+\,\mu} \bar{\sigma}_\mu \, D_L^j
+\frac{g_R}{\sqrt{2}} (V_L^*)_{ij} \, e^{i(\phi_i-\psi_j)} \, U_R^{i\,\dagger} \, W_R^{+\,\mu} \sigma_\mu \, D_R^j
+{\rm H.c.}
\nonumber \\
&=\frac{1}{\sqrt{2}} \, \bar{U}^i \, W^{+\,\mu}\gamma_\mu
\left\{g_L(V_L)_{ij}\cos\zeta P_L+g_R(V_L^*)_{ij}\,e^{i(\phi_i-\psi_j+\alpha)}\sin\zeta P_R\right\}\,D^j
\nonumber \\
&+\frac{1}{\sqrt{2}} \, \bar{U}^i \, W'^{+\,\mu} \gamma_\mu
\left\{-g_L(V_L)_{ij} \, e^{-i\,\alpha} \sin\zeta P_L+g_R(V_L^*)_{ij} \, e^{i(\phi_i-\psi_j)}\cos\zeta P_R\right\}\,D^j+{\rm H.c.},
\label{deltaf=1}
\end{align}
 where $U^i$ and $D^i$ denote the Dirac fields of the up and down-type quarks, respectively.

In this paper, we adopt the following convention for the quark phases and $\phi_u,\phi_c,\phi_t,\psi_d,\psi_s,\psi_b$:
First, we redefine the phases of five quarks to render the CKM matrix in the standard form,
\begin{align} % requires amsmath; align* for no eq. number
V_L&=   \begin{pmatrix} % or pmatrix or bmatrix or Bmatrix or ...
      c_{12}c_{13} & s_{12}c_{13} & s_{13}e^{-i\,\delta} \\
      -s_{12}c_{23}-c_{12}s_{23}s_{13}e^{i\,\delta} & c_{12}c_{23}-s_{12}s_{23}s_{13}e^{i\,\delta} & s_{23}c_{13}\\
      s_{12}s_{23}-c_{12}c_{23}s_{13}e^{i\,\delta} & -c_{12}s_{23}-s_{12}c_{23}s_{13}e^{i\,\delta} & c_{23}c_{13}\\
   \end{pmatrix}.
\label{ckm}
\end{align}
Next, we redefine $\phi_c,\phi_t,\psi_d,\psi_s,\psi_b$ to set
\begin{align} % requires amsmath; align* for no eq. number
\phi_u&=0.
\label{phiu}
\end{align}
Phase convention fixed in this way, all sources of CP violation are parametrized by
Im$[(V_L)_{cd}]$, Im$[(V_L)_{cs}]$, Im$[(V_L)_{td}]$, Im$[(V_L)_{ts}]$, Im$[(V_L)_{cd}]$,
 the newly-defined $\phi_c,\phi_t,\psi_d,\psi_s,\psi_b$, and $\alpha$.
\\

\section{$\epsilon'/\epsilon$}
\label{Sec:epsilonprime}
\subsection{Wilson Coefficients for $\Delta S=1$ Operators}

We match the $SU(3)_C\times SU(2)_L\times SU(2)_R\times U(1)_{B-L}$ gauge theory with charge symmetry
 to the effective QCD$\times$QED theory in which $W,W'$ bosons and the top quark are integrated out.
In the effective theory, the $\Delta S=1$ Hamiltonian is parametrized as
\begin{align} % requires amsmath; align* for no eq. number
{\cal H}_{\Delta S=1}&=\frac{G_F}{\sqrt{2}}\left\{
\sum_{i=1,2,...,10,1c,2c}\left(\,C_i O_i+C'_i O'_i\,\right)+\sum_{j=1,2,1c,2c}\left(\,C_j^{RL}O_j^{RL}+C_j^{LR}O_j^{LR}\,\right)+
\sum_{k=g, \gamma}
(C_k O_k+C_k^\prime O_k^\prime)
\right\}
\nonumber \\
&+{\rm H.c.},
\label{deltas=1}
\end{align}
 where operators $O$'s are defined in Appendix A.
We determine the Wilson coefficients as follows:
We approximate $g_R=g_L$ by ignoring difference in RG evolutions of $g_L$ and $g_R$ at scales below $M_{W'}$.
Also, for each Wilson coefficient, if multiple terms have an identical phase, we only consider the one in the leading order of $M_W^2/M_{W'}^2$ or $\sin\zeta$.
By integrating out $W'$, one obtains the following leading-order matching conditions at a scale $\mu\sim M_{W'}$ (note our convention with $\phi_u=0$):
\begin{align} % requires amsmath; align* for no eq. number
C'_2&=\frac{M_W^2}{M_{W'}^2}(V_L)_{us}(V_L^*)_{ud} \ e^{i(\psi_s-\psi_d)}\ \cos^2\zeta,
\label{c2p}\\
C'_{2c}&=\frac{M_W^2}{M_{W'}^2}(V_L)_{cs}(V_L^*)_{cd} \ e^{i(\psi_s-\psi_d)}\ \cos^2\zeta,
\label{c2cp}\\
C'_4&=C'_6=\frac{M_W^2}{M_{W'}^2}\frac{\alpha_s}{4\pi}\cos^2\zeta\sum_{i=u,c,t}(V_L)_{is}(V_L^*)_{id} \ e^{i(\psi_s-\psi_d)}
\ \frac{1}{2}F_1(y_i),
\ \ \ \ \
C'_3=C'_5=-\frac{1}{3}C'_4,
\label{cp3456}\\
C'_7&=C'_9=\frac{M_W^2}{M_{W'}^2}\frac{\alpha}{4\pi}\cos^2\zeta\sum_{i=u,c,t}(V_L)_{is}(V_L^*)_{id} \ e^{i(\psi_s-\psi_d)}\ \frac{2}{3}E_{1d}(y_i),
\label{cp79}\\
\delta C_g&=\frac{M_W^2}{M_{W'}^2}\cos^2\zeta\frac{m_d}{m_s}\sum_{i=u,c,t}(V_L)_{is}(V_L^*)_{id} \ F_2(y_i),
\\
\delta C_\gamma&=\frac{M_W^2}{M_{W'}^2}\cos^2\zeta\frac{m_d}{m_s}\sum_{i=u,c,t}(V_L)_{is}(V_L^*)_{id} \ E_{2d}(y_i),
\\
\delta C'_g&=\frac{M_W^2}{M_{W'}^2}\cos^2\zeta\sum_{i=u,c,t}(V_L)_{is}(V_L^*)_{id} \ F_2(y_i),
\\
\delta C'_\gamma&=\frac{M_W^2}{M_{W'}^2}\cos^2\zeta\sum_{i=u,c,t}(V_L)_{is}(V_L^*)_{id} \ E_{2d}(y_i),\label{Eq:Wp}
\\
&{\rm with} \ y_i\equiv m_i^2/M_{W'}^2.
\nonumber
\end{align}
By further integrating out $W$ and the top quark, one gains the following leading-order matching conditions at a scale $\mu\sim M_W$ (note our convention with $\phi_u=0$):
\begin{align} % requires amsmath; align* for no eq. number
C_2&=(V_L^*)_{us}(V_L)_{ud}\,\cos^2\zeta,
\label{c2}\\
C_{2c}&=(V_L^*)_{cs}(V_L)_{cd}\,\cos^2\zeta,
\label{c2c}\\
C_2^{RL}&=(V_L)_{us}(V_L)_{ud} \ e^{i(\psi_s-\alpha)}\,\sin\zeta\cos\zeta, \ \ \
C_2^{LR}=(V_L^*)_{us}(V_L^*)_{ud} \ e^{i(-\psi_d+\alpha)}\,\sin\zeta\cos\zeta,
\label{c2lr}\\
C_{2c}^{RL}&=(V_L)_{cs}(V_L)_{cd} \ e^{i(-\phi_c+\psi_s-\alpha)}\,\sin\zeta\cos\zeta, \ \ \
C_{2c}^{LR}=(V_L^*)_{cs}(V_L^*)_{cd} \ e^{i(\phi_c-\psi_d+\alpha)}\,\sin\zeta\cos\zeta,
\\
C_4&=C_6=\frac{\alpha_s}{4\pi}\cos^2\zeta\sum_{i=u,c,t}(V_L^*)_{is}(V_L)_{id}\ \frac{1}{2}F_1(x_i), \ \ \ \ \ C_3=C_5=-\frac{1}{3}C_4,
\\
C_7&=C_9=\frac{\alpha}{4\pi}\cos^2\zeta\sum_{i=u,c,t}(V_L^*)_{is}(V_L)_{id}\ \frac{2}{3}E_{1d}(x_i),
\label{c7}\\
\Delta C_g&=\sum_{i=u,c,t}\left\{\cos^2\zeta(V_L^*)_{is}(V_L)_{id} \ F_2(x_i)
+\sin\zeta\cos\zeta\frac{m_i}{m_s}(V_L)_{is}(V_L)_{id} \ e^{i(-\phi_i+\psi_s-\alpha)}F_3(x_i)\right\},
\\
\Delta C_\gamma&=\sum_{i=u,c,t}\left\{\cos^2\zeta(V_L^*)_{is}(V_L)_{id} \ E_{2d}(x_i)
+\sin\zeta\cos\zeta\frac{m_i}{m_s}(V_L)_{is}(V_L)_{id} \ e^{i(-\phi_i+\psi_s-\alpha)}E_{3d}(x_i)\right\},
\\
\Delta C'_g&=\sum_{i=u,c,t}\left\{\cos^2\zeta\frac{m_d}{m_s}(V_L^*)_{is}(V_L)_{id} \ F_2(x_i)
+\sin\zeta\cos\zeta\frac{m_i}{m_s}(V_L^*)_{is}(V_L^*)_{id} \ e^{i(\phi_i-\psi_d+\alpha)}F_3(x_i)\right\},
\\
\Delta C'_\gamma&=\sum_{i=u,c,t}\left\{\cos^2\zeta\frac{m_d}{m_s}(V_L^*)_{is}(V_L)_{id} \ E_{2d}(x_i)
+\sin\zeta\cos\zeta\frac{m_i}{m_s}(V_L^*)_{is}(V_L^*)_{id} \ e^{i(\phi_i-\psi_d+\alpha)}E_{3d}(x_i)\right\},
\label{cpg}\\
&{\rm with} \ x_i\equiv m_i^2/M_W^2,
\nonumber
\end{align}
 where loop functions $F_1,F_2,F_3$ and $E_{1d},E_{2d},E_{3d}$ are defined in Appendix~B.
We are aware that the dipole operators receive two contributions with different phases when
 $W'$ is integrated out and when $W$ is.
The two are expressed as $\delta C_g,\delta C_\gamma,\delta C'_g,\delta C'_\gamma$
 and $\Delta C_g,\Delta C_\gamma,\Delta C'_g,\Delta C'_\gamma$, respectively.
\\

We take into account RG evolutions of the Wilson coefficients at order $O(\alpha_s)$.
The fact that four sets of operators, $(\{O_i\})$, $(\{O'_i\})$ $(i=1,2,...,10,1c,2c)$, $(\{O^{RL}_j\})$, $(\{O^{LR}_j\})$ $(j=1,2,1c,2c)$,
 do not mix with each other facilitates the computation.
For $(\{C_i\})$, $(\{C^{RL}_j\})$ and $(\{C^{LR}_j\})$, we assume that their initial conditions at scale $\mu=M_W$ are given by
 Eqs.~(\ref{c2}--\ref{c7}) and solve the RG equations from $\mu=M_W$ to the scale for which the lattice results are reported.
For $(\{C'_i\})$, we assume that their initial conditions at $\mu=M_{W'}$ are provided by Eqs.~(\ref{c2p}-\ref{Eq:Wp})
and solve the RG equations from $\mu=M_{W'}$ to the scale of lattice results.
Finally, we compute RG evolutions of the coefficients of the dipole operators $(C_g,C_\gamma)$ and $(C'_g,C'_\gamma)$,
 which receive contributions from $(\{C_i,C^{RL}_j\})$ and $(\{C'_i,C^{LR}_j\})$, respectively.
The $O(\alpha_s)$ RG equations for $(\{C_i\})$ and $(\{C^{RL}_i\})$ are found in Ref.~\cite{Ciuchini},
and those for $(\{C^{RL}_i\})$ and
$(C_g,C_\gamma)$ are in Ref.~\cite{Cho:1993zb}.
\\

\subsection{Hadronic Matrix Elements}

We employ the lattice calculations of hadronic matrix elements 
$\langle (\pi\pi)_I \vert O_i \vert K^0\rangle$
for $i=1,2,...,10$ for $I=0,2$ reported by 
RBC/UKQCD in Refs.~\cite{Blum:2015ywa, Bai:2015nea}.

Since lattice calculations for the matrix elements of $O_1^{LR}$ and $O_2^{LR}$ are missing, we estimate them from the RBC/UKQCD results using isospin symmetry.
In the limit of exact isospin symmetry, we find, for $\Delta I=3/2$ amplitudes,
\begin{align} % requires amsmath; align* for no eq. number
\langle (\pi\pi)_{I=2} \vert O_7 \vert K^0\rangle&=
\langle (\pi\pi)_{I=2} \vert (\bar{s}d)_L(\bar{u}u-\frac{1}{2}\bar{d}d-\frac{1}{2}\bar{s}s)_R \vert K^0\rangle
\nonumber \\
&=\frac{3}{4}\langle (\pi\pi)_{I=2} \vert (\bar{s}d)_L(\bar{u}u-\bar{d}d)_R \vert K^0\rangle
\\
&=\frac{3}{4}\langle (\pi\pi)_{I=2} \vert \sqrt{2}(\bar{s}u)_L\sqrt{2}(\bar{u}d)_R \vert K^0\rangle
\\
&=\frac{3}{2}\langle (\pi\pi)_{I=2} \vert O_2^{LR} \vert K^0\rangle,
\end{align}
 where we have discarded $\Delta I=1/2$ part when obtaining the second line, 
 and when deriving the third line, we have inserted Clebsch-Gordan coefficients for constructing the $\Delta I=3/2$
 operator from a $\Delta I=1/2$ one and a $\Delta I=1$ one.
For $\Delta I=1/2$ amplitudes, we find
\begin{align} % requires amsmath; align* for no eq. number
\langle (\pi\pi)_{I=0} \vert (\frac{4}{3}O_7+\frac{2}{3}O_5) \vert K^0\rangle
&=\langle (\pi\pi)_{I=0} \vert
\left\{(\bar{s}d)_L(\bar{u}u+\bar{d}d)_R+(\bar{s}d)_L(\bar{u}u-\bar{d}d)_R\right\}
\vert K^0\rangle
\\
&=\langle (\pi\pi)_{I=0} \vert \{
\sqrt{\frac{3}{2}}(\bar{s}u)_L\sqrt{2}(\bar{u}d)_R 
-\sqrt{\frac{1}{3}}\sqrt{\frac{3}{2}}(\bar{s}u)_L\sqrt{2}(\bar{u}d)_R \}\vert K^0\rangle
\\
&=(\sqrt{3}-1)\langle (\pi\pi)_{I=0} \vert O_2^{LR}\vert K^0\rangle,
\end{align}
 where in the first line, we have separated $(\bar{u}u)_R$ into $\Delta I=0$ and $\Delta I=1$ parts,
 and in the second line, we have inserted Clebsch-Gordan coefficients for constructing the $\Delta I=1/2$ operator from 
 a $\Delta I=1/2$ one and a $\Delta I=0$ one or a $\Delta I=1$ one.
The same relations hold between the matrix elements for $O_1^{LR}$ and $O_8,O_6$.

The hadronic matrix elements for the chromo-dipole operators $O_g,O'_g$ are extracted from the calculation based on dual QCD approach~\cite{Buras:2018evv}.
Note that the above calculation is corroborated by the fact that it is consistent
 with a lattice calculation of the $K$-$\pi$ hadronic matrix element~\cite{Constantinou:2017sgv}, which is related to the $K$-$\pi\pi$ one by chiral perturbation theory.

\subsection{Numerical Analysis of $\epsilon'/\epsilon$}

The definition for the decay amplitudes of $K^0\to\pi\pi$ is
\begin{eqnarray}
A_0e^{i\delta_0}=\bra{(\pi\pi)_{\mathrm{I}=0}}\mathcal{H}_{\Delta S=1}\ket{K^0},\quad
A_2e^{i\delta_2}=\bra{(\pi\pi)_{\mathrm{I}=2}}\mathcal{H}_{\Delta S=1}\ket{K^0},
\end{eqnarray}
where $\delta_{0, 2}$ represent the strong phases.
In terms of the above amplitudes, one writes
the direct CP violation parameter divided by the indirect one as
\label{Sec:epsilon}
\begin{eqnarray}
\mathrm{Re}\left(\frac{\epsilon^\prime}{\epsilon}\right)
=\mathrm{Re}\left(\frac{i\omega e^{i(\delta_2-\delta_0)}}{\sqrt{2}}\right)\left(\frac{\mathrm{Im} A_2}{\mathrm{Re}A_2}-\frac{\mathrm{Im} A_0}{\mathrm{Re}A_0}\right),
\end{eqnarray}
where $\omega=\mathrm{Re} A_2/\mathrm{Re} A_0$ is a suppression factor due to the $\Delta I=1/2$ rule.
For the strong phases,
we use the values of Refs.~\cite{Bai:2015nea,Blum:2015ywa},
$\delta_2=23.8\pm5.0\ \mathrm{degree}$ and $\delta_0=-11.6\pm2.8\ \mathrm{degree}$.
For the real parts of the decay amplitudes,
we employ the experimental data \cite{Patrignani:2016xqp},
$\mathrm{Re}A_2=1.479\times 10^{-8}\ \mathrm{GeV}$ and
$\mathrm{Re}A_0=33.20\times 10^{-8}\ \mathrm{GeV}$,
which leads to $\omega=4.454\times 10^{-2}$.
In our analysis, we separate the SM and new physics contributions as
\begin{eqnarray}
\mathrm{Re}\left(\frac{\epsilon^\prime}{\epsilon}\right)
&=&\mathrm{Re}\left(\frac{\epsilon^\prime}{\epsilon}\right)_{\mathrm{SM}}+
\mathrm{Re}\left(\frac{\epsilon^\prime}{\epsilon}\right)_{\mathrm{NP}}.\label{Eq:DefEps}
\end{eqnarray}
For the SM part, we quote the calculation in the literature
$\mathrm{Re}(\epsilon^\prime/\epsilon)_{\rm SM}=(1.38\pm6.90)
\times10^{-4}$ \cite{Bai:2015nea}.
It is the new physics part,
\begin{eqnarray}
\mathrm{Re}\left(\frac{\epsilon^\prime}{\epsilon}\right)
_{\mathrm{NP}}&=&
\mathrm{Re}\left(\frac{i\omega e^{i(\delta_2-\delta_0)}}{\sqrt{2}}\right)\left(\frac{\mathrm{Im} A_2^{\mathrm{NP}}}{\mathrm{Re}A_2}-\frac{\mathrm{Im} A_0^{\mathrm{NP}}}{\mathrm{Re}A_0}\right),\label{Eq:Amp}
\end{eqnarray}
 that we compute in this paper.
In doing so, we approximate $\cos^2\zeta=1$ in the Wilson coefficients Eqs.~(\ref{c2}--\ref{cpg}),
 so that the SM contribution is separated from the new physics one at the operator level.
\\

In the analysis, we fix the ratio of the bifundamental scalar VEVs at its natural value as $\tan\beta=m_b/m_t$.
We have found numerically that for $M_{W^\prime}> 1\ \mathrm{TeV}$,
the chromo-dipole contribution to $\mathrm{Re}(\epsilon^\prime/\epsilon)$ does not exceed $\mathcal{O}(10^{-4})$ and is hence safely neglected
\footnote{
When we use a calculation based on the chiral quark model in Ref.~\cite{Bertolini:2012pu} to evaluate the hadronic matrix elements of the chromo-dipole operators,
 we are again lead to the result that the chromo-dipole contribution to $\mathrm{Re}(\epsilon^\prime/\epsilon)$ is below $\mathcal{O}(10^{-4})$ for $M_{W^\prime}> 1\ \mathrm{TeV}$.
}.
Consequently, only two combinations of new CP phases, $\alpha-\psi_d$ and $\alpha-\psi_s$, and the $W'$ mass determine
the new physics contribution.

\begin{figure}[H]
\centering
\includegraphics[width=12.0cm]{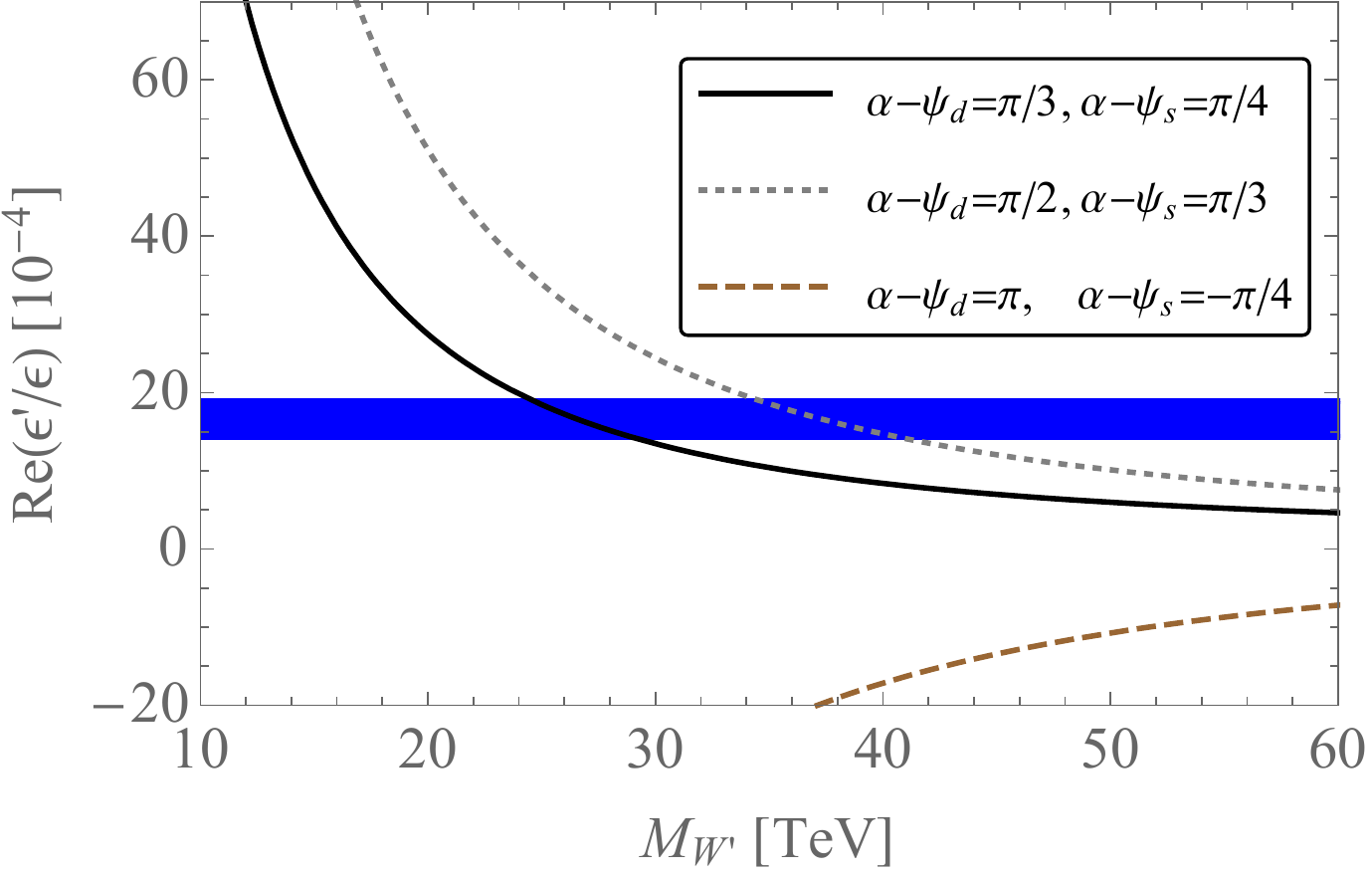}
  \caption{Numerical result of $\mathrm{Re}(\epsilon^\prime/\epsilon)$.
  The blue band represents the 1$\sigma$ range
  of experimental data given by PDG \cite{Patrignani:2016xqp},
  while model predictions
  with specific choice of phases are shown by
  the lines.}
\label{Fig:1}
\end{figure}

First, we choose specific values for the new CP phases
in the calculation of $\epsilon^\prime/\epsilon$
to illustrate the model prediction.
In Fig.\ \ref{Fig:1}, the prediction for
$\mathrm{Re}(\epsilon^\prime/\epsilon)$
is presented with specific choices of
$\alpha-\psi_d$ and $\alpha-\psi_s$.
\par
\begin{figure}[H]
\centering
\includegraphics[width=12.0cm]{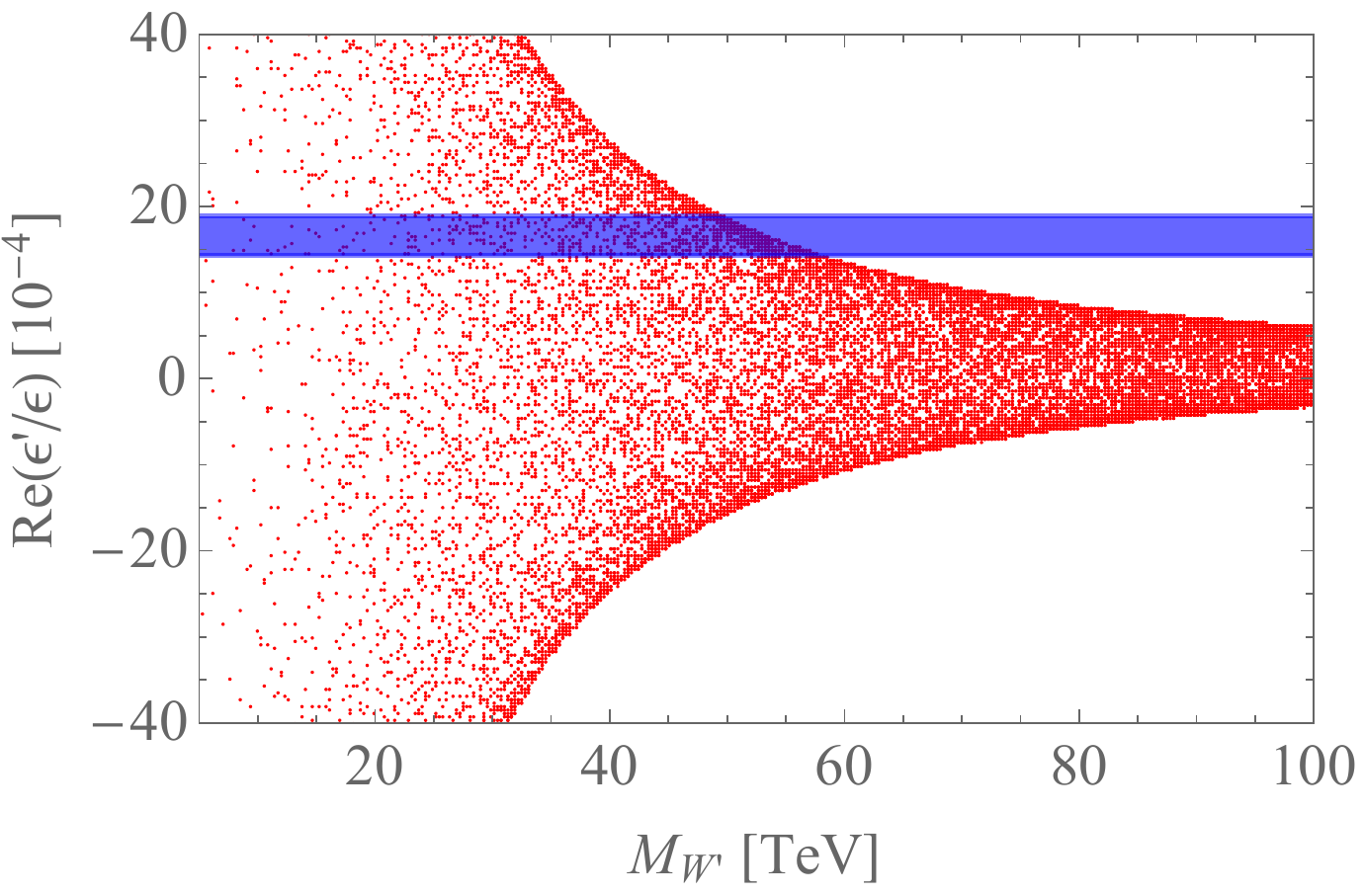}
  \caption{
  Numerical result of $\mathrm{Re}(\epsilon^\prime/\epsilon)$.
  The blue band represents the experimental
  data given by PDG \cite{Patrignani:2016xqp}, while each red dot corresponds to
  the model prediction with a randomly generated set of ($\alpha-\psi_d$, $\alpha-\psi_s$).}
\label{Fig:2}
\end{figure}

Next, we randomly vary $\alpha-\psi_d$ and $\alpha-\psi_s$ in the range
$[0, 2\pi]$, since they are free parameters.
In Fig.\ \ref{Fig:2}, we show the region of
$\mathrm{Re}(\epsilon^\prime/\epsilon)$
obtained by varying $\alpha-\psi_d$ and
$\alpha-\psi_s$.
One observes that $M_{W^\prime}<58\ \mathrm{TeV}$
is necessary for $1\sigma$ explanation of
the anomaly.
We have confirmed that among the terms of $\Delta S=1$ Hamiltonian Eq.~(\ref{deltas=1}),
$\sum_{i=1,2}C'_iO'_i$ and $\sum_{j=1,2}(C_j^{RL}O_j^{RL}+C_j^{LR}O_j^{LR})$ are the leading sources of the new physics contribution.

\section{Neutron Electric Dipole Moment}
\label{Sec:EDM}
\subsection{Wilson Coefficients for Operators contributing to the neutron EDM}

In the effective QCD$\times$QED theory in which $W,W'$ and the top quark are integrated out,
 the part of the CP-violating Hamiltonian that contributes to the neutron EDM is parametrized as
\begin{align} % requires amsmath; align* for no eq. number
{\cal H}_{nEDM}&=\frac{G_F}{\sqrt{2}}\left\{
\sum_{q}\sum_{i=1,2,4,5}{\cal C}_{iq} {\cal O}_{iq}+{\cal C}_3 {\cal O}_3+\sum_{q\neq q'}\sum_{i=1}^4{\cal C}_{iq'q} {\cal O}_{iq'q}
\right\},
\label{cpv}
\end{align}
 where operators ${\cal O}$'s are defined in Appendix C.

We determine the Wilson coefficients ${\cal C}$'s as follows:
Again, for each coefficient, if multiple terms have an identical phase, we exclusively consider the one in the leading order of $M_W^2/M_{W'}^2$ or $\sin\zeta$.
By integrating out $W$ and the top quark, one obtains the following leading-order matching conditions at $\mu\sim M_W$ (note our convention with $\phi_u=0$):
\begin{align} % requires amsmath; align* for no eq. number
{\cal C}_{2du}&=-{\cal C}_{2ud}=4\sin\zeta\cos\zeta \ {\rm Im}\left[(V_L)_{ud}(V_L)_{ud}\ e^{i(\psi_d-\alpha)}\right],
\ \ \ \ \ (d\to s),
\label{edm2}\\
{\cal C}_{2dc}&=-{\cal C}_{2cd}=4\sin\zeta\cos\zeta \ {\rm Im}\left[(V_L)_{cd}(V_L)_{cd}\ e^{i(-\phi_c+\psi_d-\alpha)}\right],
\ \ \ \ \ (d\to s),
\\
{\cal C}_{1u}&=\frac{e}{4\pi^2}\sin\zeta\cos\zeta\sum_{i=d,s,b}\frac{m_i}{m_u}\,
{\rm Im}\left[(V_L)_{ui}(V_L)_{ui}\ e^{i(\psi_i-\alpha)}\right]\,E_{3u}(x_i),
\\
{\cal C}_{1c}&=\frac{e}{4\pi^2}\sin\zeta\cos\zeta\sum_{i=d,s,b}\frac{m_i}{m_u}\,
{\rm Im}\left[(V_L)_{ui}(V_L)_{ui}\ e^{i(-\phi_c+\psi_i-\alpha)}\right]\,E_{3u}(x_i),
\\
{\cal C}_{1d}&=\frac{e}{4\pi^2}\sin\zeta\cos\zeta\sum_{i=u,c,t}\frac{m_i}{m_d}\,
{\rm Im}\left[(V_L)_{id}(V_L)_{id}\ e^{i(-\phi_i+\psi_d-\alpha)}\right]\,E_{3d}(x_i), \ \ \ (d \to s),
\\
{\cal C}_{2u}&=-\frac{g_s}{4\pi^2}\sin\zeta\cos\zeta\sum_{i=d,s,b}\frac{m_i}{m_u}\,
{\rm Im}\left[(V_L)_{ui}(V_L)_{ui}\ e^{i(\psi_i-\alpha)}\right]\,F_3(x_i),
\\
{\cal C}_{2c}&=-\frac{g_s}{4\pi^2}\sin\zeta\cos\zeta\sum_{i=d,s,b}\frac{m_i}{m_u}\,
{\rm Im}\left[(V_L)_{ui}(V_L)_{ui}\ e^{i(-\phi_c+\psi_i-\alpha)}\right]\,F_3(x_i),
\\
{\cal C}_{2d}&=\frac{g_s}{4\pi^2}\sin\zeta\cos\zeta\sum_{i=u,c,t}\frac{m_i}{m_d}\,
{\rm Im}\left[(V_L)_{id}(V_L)_{id}\ e^{i(-\phi_i+\psi_d-\alpha)}\right]\,F_3(x_i), \ \ \ (d \to s),
\\
{\cal C}_3&\simeq\frac{4g_s^3}{(16\pi^2)^2}\sin\zeta\cos\zeta \ \frac{m_t}{m_b}F_3(x_t) \,
{\rm Im}\left[(V_L)_{tb}(V_L)_{tb}\ e^{i(-\phi_t+\psi_b-\alpha)}\right],
\label{edmw}
\\
&{\rm with} \ x_i\equiv m_i^2/M_W^2,
\nonumber
\end{align}
 where loop functions $F_1,F_2,F_3$ and $E_{1d},E_{2d},E_{3d},E_{3u}$ are defined in Appendix~B.
In Eq.~(\ref{edmw}) (which corresponds to the Weinberg operator~\cite{Weinberg:1989dx}), 
 we present the dominant part proportional to $m_t$.
Terms obtained by integrating out $W'$ possess the same phases as Eqs.~(\ref{edm2}--\ref{edmw}) and are simply suppressed by $M_W^2/M_{W'}^2$ compared to Eqs.~(\ref{edm2}--\ref{edmw}).
They are therefore neglected in our analysis.
\\

The RG equations at order $O(\alpha_s)$ for the Wilson coefficients are obtainable in Refs. \cite{Shifman:1976de,
Dai:1989yh,
Boyd:1990bx, 
Hisano:2012cc}.
We assume that the initial conditions at $\mu=M_W$ for the RG equations are given by Eqs.~(\ref{edm2}--\ref{edmw}), 
 and solve the equations from $\mu=M_W$ to $\mu=1$~GeV.
At the 1~GeV scale, we evaluate the hadronic matrix elements.

\subsection{Hadronic Matrix Elements}

\subsubsection{Four-quark operators ${\cal O}_{1q'q}$, ${\cal O}_{2q'q}$}

In the $SU(2)_L\times SU(2)_R\times U(1)_{B-L}$ model with charge symmetry,
 the Wilson coefficients for the four-quark operators ${\cal O}_{1q'q}=(\bar{q}'q')(\bar{q}i\gamma_5q)$ and ${\cal O}_{2q'q}=(\bar{q}'_\alpha q'_\beta)(\bar{q}_\beta i\gamma_5q_\alpha)$ ($q\neq q';\ q,q'=u,d,s$)
 are particularly large.
Therefore, we scrutinize how these operators contribute to the neutron EDM.
Operators ${\cal O}_{1q'q}$ contribute in the following three ways:
\begin{itemize}
\item The first one is through meson condensation~\cite{deVries:2012ab}; ${\cal O}_{1q'q}$ operators give rise to tadpole terms for pseudoscalar mesons and induce their VEVs. 
These VEVs generate CP-violating interactions for baryons and mesons, which
 contribute to the neutron EDM through baryon-meson loop diagrams.

\item The second one is through hadronic matrix elements of ${\cal O}_{1q'q}$ with baryons and mesons, 
 $\langle B M \vert{\cal O}_{1q'q}\vert B\rangle$ ($B$ denotes a baryon and $M$ a meson), which contribute to 
 the neutron EDM through baryon-meson loop diagrams.

\item The third one is directly through the hadronic matrix element of ${\cal O}_{1q'q}$ with neutrons and photon.
\end{itemize}
On the other hand,  ${\cal O}_{2q'q}$ operators do not yield meson condensation, but do contribute to the neutron EDM in the latter two ways.
Later, it will be shown that the contribution from the pion VEV $\langle\pi^0\rangle$, which belongs to the first category, is enhanced by the factor $m_s/(m_u+m_d)$ compared to the latter two.
We therefore investigate how ${\cal O}_{1q'q}$ operators bring about meson condensation, thereby contributing to the neutron EDM.

We are aware that if Peccei-Quinn mechanism~\cite{pq} exists, it affects the meson condensation
 and also induces an effective non-zero $\bar{\theta}$ term due to incomplete cancellation between the genuine $\bar{\theta}$ term and the axion VEV.
Alternatively, it is logically possible to assume $\bar{\theta}=0$ without Peccei-Quinn mechanism,
 by considering an unknown mechanism or through fine-tuning,
 in which case we do not need to take into account 
 the effect of Peccei-Quinn mechanism or that of non-zero $\bar{\theta}$.
In this paper, we consider both cases where (i) one has $\bar{\theta}=0$ without Peccei-Quinn mechanism, and (ii) Peccei-Quinn mechanism is at work.
\\

We start from the case with $\bar{\theta}=0$ without Peccei-Quinn mechanism.
The meson condensation contribution is evaluated by the following steps:
\\

\noindent
(1) First, we implement $\sum{\cal C}_{1q'q}{\cal O}_{1q'q}$ part of the Hamiltonian Eq.~(\ref{cpv})
 into the meson chiral Lagrangian.
To this end, we rewrite
\begin{align} % requires amsmath; align* for no eq. number
\sum_{q\neq q';q,q'=u,d,s}{\cal C}_{1q'q}{\cal O}_{1q'q}&=\sum_{i,j,k,l=u,d,s}\left\{ i{\cal C}_{ijkl}^{LRLR} \ (\bar{q}_{iL}q_{jR})(\bar{q}_{kL}q_{lR})
+i{\cal C}_{ijkl}^{RLLR} \ (\bar{q}_{iR}q_{jL})(\bar{q}_{kL}q_{lR}) \right\}-(L\leftrightarrow R),
\\
{\rm with}\ \ {\cal C}_{ijkl}^{LRLR}&={\cal C}_{ijkl}^{RLLR}\equiv\sum_{q\neq q'}{\cal C}_{1q'q} \ \delta_{iq'}\delta_{jq'}\delta_{kq}\delta_{lq}.
\label{csdef}
\end{align}
It then becomes clear that the theory would be invariant (except for $U(1)_A$ anomaly) if coefficients ${\cal C}_{ijkl}^{LRLR}$ and ${\cal C}_{ijkl}^{RLLR}$ transformed under $U(3)_L\times U(3)_R$ rotations, $L\times R$, as 
\begin{align*} % requires amsmath; align* for no eq. number
{\cal C}_{ijkl}^{LRLR}&\to\sum_{m,n,o,p}(L)_{im}(L)_{ko} \ {\cal C}_{mnop}^{LRLR} \ (R^\dagger)_{nj}(R^\dagger)_{pl},
\\
{\cal C}_{ijkl}^{RLLR}&\to\sum_{m,n,o,p}(R)_{im}(L)_{ko} \ {\cal C}_{mnop}^{RLLR} \ (L^\dagger)_{nj}(R^\dagger)_{pl}.
\end{align*}
From the above transformation property and the parity invariance of QCD, 
 the meson chiral Lagrangian at order $O(p^2)$ plus the leading CP-violating terms is found to be
 (remind that $\bar{\theta}=0$ has been assumed)
\begin{align} % requires amsmath; align* for no eq. number
{\cal L}_{\rm mesons}&=\frac{F_\pi^2}{4}{\rm tr}\left[(D_\mu U)^\dagger D^\mu U+\chi(U+U^\dagger)\right]
+\frac{F_0^2-F_\pi^2}{12}{\rm tr}\left[UD_\mu U^\dagger\right]{\rm tr}\left[U^\dagger D^\mu U\right]
\nonumber\\
&+a_0 \, {\rm tr}\left[\log U-\log U^\dagger\right]^2
\nonumber \\
&+\frac{G_F}{\sqrt{2}}\sum_{i,j,k,l=u,d,s}\left\{
i{\cal C}_{ijkl}^{LRLR} \left(c_1 [U]_{ji}[U]_{lk}-c_1 [U^\dagger]_{ji}[U^\dagger]_{lk}+c_2 [U]_{li}[U]_{jk}-c_2 [U^\dagger]_{li}[U^\dagger]_{jk}\right)\right.
\nonumber \\
&\left.+i{\cal C}_{ijkl}^{RLLR} \left(c_3 [U^\dagger]_{ji}[U]_{lk}-c_3 [U]_{ji}[U^\dagger]_{lk}\right)\right\},
\label{meson}
\end{align}
 where ${\cal C}_{ijkl}^{LRLR},\,{\cal C}_{ijkl}^{RLLR}$ have been defined in Eq.~(\ref{csdef}).
Here, $U$ is a nonlinear representation of the nine Nambu-Goldstone bosons that transforms under $U(3)_L\times U(3)_R$ rotations $L\times R$ as $U\to RUL^\dagger$,
and $\chi$ includes the quark mass term, which are given by
\begin{align}
U&=\exp\left[\frac{2i}{\sqrt{6}F_0}\eta_0\,I_3+\frac{2i}{F_\pi}\Pi\right],
\ \ \ \ \
\Pi\equiv\begin{pmatrix} % or pmatrix or bmatrix or Bmatrix or ...
      \frac{1}{2}\pi^0+\frac{1}{2\sqrt{3}}\eta_8 & \frac{1}{\sqrt{2}}\pi^+ & \frac{1}{\sqrt{2}}K^+ \\
     \frac{1}{\sqrt{2}}\pi^-  &  -\frac{1}{2}\pi^0+\frac{1}{2\sqrt{3}}\eta_8 & \frac{1}{\sqrt{2}}K^0 \\
     \frac{1}{\sqrt{2}}K^- & \frac{1}{\sqrt{2}}\bar{K}^0 & -\frac{1}{\sqrt{3}}\eta_8 \\
   \end{pmatrix},
\\
I_3&\equiv{\rm diag}(1,~1,~1),
\nonumber\\
\chi&=2B_0\ {\rm diag}(m_u,~m_d,~m_s).
\end{align}
$[U]_{ij}$ denotes the $(i,j)$ component of matrix $U$.
$F_\pi$ is the pion decay constant in the chiral limit and $F_0$ is the decay constant for $\eta_0$, which we approximate as $F_0\simeq F_\pi$.
$B_0$ satisfies $B_0\simeq m_\pi^2/(m_u+m_d)$.
The term with $\log U$ represents instanton effects, whose expression is exact in the large $N_c$ limit~\cite{wv},
 and $a_0$ satisfies $48a_0/F_0^2\simeq m_\eta^2+m_{\eta'}^2-2m_K^2$.
$c_1$, $c_2$ and $c_3$ are unknown low energy constants (LECs),
 which can be estimated by na\"ive dimensional analysis~\cite{manohar} as
\begin{align} % requires amsmath; align* for no eq. number
c_1&\sim c_2\sim c_3\sim \frac{(4\pi F_\pi)^6}{(4\pi)^4}.
\label{c1c2c3}
\end{align}
\\

\noindent
(2) The CP-violating part of the Lagrangian Eq.~(\ref{meson}) contains tadpole terms for mesons, which lead to non-zero meson VEVs.
Assuming that electric charge and strangeness are not broken spontaneously, we obtain the following potential for neutral mesons $\pi^0$, $\eta_8$ and $\eta_0$:
\begin{align} % requires amsmath; align* for no eq. number
&V(\pi^0,\eta_8,\eta_0)=F_\pi^2B_0\left\{
m_u\cos\left(\frac{\pi^0}{F_\pi}+\frac{\eta_8}{\sqrt{3}F_\pi}+\frac{2\eta_0}{\sqrt{6}F_0}\right)
+m_d\cos\left(-\frac{\pi^0}{F_\pi}+\frac{\eta_8}{\sqrt{3}F_\pi}+\frac{2\eta_0}{\sqrt{6}F_0}\right)\right.
\nonumber \\
&\left.+m_s\cos\left(-\frac{2\eta_8}{\sqrt{3}F_\pi}+\frac{2\eta_0}{\sqrt{6}F_0}\right) \right\}-24\frac{a_0}{F_0^2}(\eta_0)^2
\nonumber\\
&-\frac{G_F}{\sqrt{2}}2c_1\left\{({\cal C}_{1ud}+{\cal C}_{1du})\sin\left(\frac{2\eta_8}{\sqrt{3}F_\pi}+\frac{4\eta_0}{\sqrt{6}F_0}\right)+
({\cal C}_{1us}+{\cal C}_{1su})\sin\left(\frac{\pi^0}{F_\pi}-\frac{\eta_8}{\sqrt{3}F_\pi}+\frac{4\eta_0}{\sqrt{6}F_0}\right)\right.
\nonumber\\
&\left.+({\cal C}_{1ds}+{\cal C}_{1sd})\sin\left(-\frac{\pi^0}{F_\pi}-\frac{\eta_8}{\sqrt{3}F_\pi}+\frac{4\eta_0}{\sqrt{6}F_0}\right)\right\}
\nonumber\\
&-\frac{G_F}{\sqrt{2}}2c_3\left\{({\cal C}_{1ud}-{\cal C}_{1du})\sin\left(-\frac{2\pi^0}{F_\pi}\right)+
({\cal C}_{1us}-{\cal C}_{1su})\sin\left(-\frac{\pi^0}{F_\pi}-\frac{\sqrt{3}\eta_8}{F_\pi}\right)\right.
\nonumber\\
&\left.+({\cal C}_{1ds}-{\cal C}_{1sd})\sin\left(\frac{\pi^0}{F_\pi}-\frac{\sqrt{3}\eta_8}{F_\pi}\right)\right\}.
\label{mesonpotential}
\end{align}
The above potential is minimized with non-zero meson VEVs, $\langle\pi^0\rangle$, $\langle\eta_8\rangle$ and $\langle\eta_0\rangle$.
Insofar as we are concerned with vertices with one meson,
 the physical modes of $\pi^0$, $\eta_8$ and $\eta_0$ fields can be approximated as
\begin{align} % requires amsmath; align* for no eq. number
\pi^0_{\rm phys}&\simeq\pi^0-\langle\pi^0\rangle, \ \ \ \eta_{\rm 8 phys}\simeq\eta_8-\langle \eta_8\rangle, \ \ \ 
\eta_{\rm 0 phys}\simeq\eta_0-\langle \eta_0\rangle.
\label{physmode}
\end{align}

In the $SU(2)_L\times SU(2)_R\times U(1)_{B-L}$ model with charge symmetry, there hold relations
 ${\cal C}_{1ud}\simeq -{\cal C}_{1du}$ and $\vert{\cal C}_{1ud}\vert \gg \vert{\cal C}_{1sq}\vert,\,\vert{\cal C}_{1qs}\vert$ $(q=u,d)$.
When $({\cal C}_{1ud}+{\cal C}_{1du})$ and ${\cal C}_{1sq},\,{\cal C}_{1qs}$ are neglected accordingly, one finds, for small VEVs,
\begin{align} % requires amsmath; align* for no eq. number
\frac{\langle\pi^0\rangle}{F_\pi}&\simeq
\frac{G_F}{\sqrt{2}}({\cal C}_{1ud}-{\cal C}_{1du})\frac{c_3}{B_0F_\pi^2}
\frac{B_0F_\pi^2(m_u+m_d)m_s+8a_0(m_u+m_d+4m_s)}{B_0F_\pi^2m_um_dm_s+8a_0(m_um_d+m_dm_s+m_sm_u)},
\nonumber\\
\frac{\langle\eta_8\rangle}{F_\pi}&\simeq
\frac{G_F}{\sqrt{2}}({\cal C}_{1ud}-{\cal C}_{1du})\frac{c_3}{\sqrt{3}B_0F_\pi^2}
(m_d-m_u)\frac{B_0F_\pi^2m_s+24a_0}{B_0F_\pi^2m_um_dm_s+8a_0(m_um_d+m_dm_s+m_sm_u)},
\nonumber \\
\frac{\langle\eta_0\rangle}{F_0}&\simeq
\frac{G_F}{\sqrt{2}}({\cal C}_{1ud}-{\cal C}_{1du})\frac{\sqrt{2}c_3}{\sqrt{3}B_0F_\pi^2}
(m_d-m_u)\frac{B_0F_\pi^2m_s}{B_0F_\pi^2m_um_dm_s+8a_0(m_um_d+m_dm_s+m_sm_u)}.
\label{vevestimate}
\end{align}
Note that $\langle\eta_8\rangle$ and $\langle\eta_0\rangle$ are proportional to $m_d-m_u$.
This is because these VEVs are isospin singlets and hence must be constructed from the product of
 isospin-odd coefficient ${\cal C}_{1ud}-{\cal C}_{1du}$ and isospin-odd mass term $m_d-m_u$.
In contrast, $\langle\pi^0\rangle$ does not contain $m_d-m_u$ because this VEV is isospin-violating.
Since $20 a_0 \sim B_0 F_\pi^2 m_s$ holds empirically,
 we find from Eq.~(\ref{vevestimate}) that $\langle\pi^0\rangle$ is much larger than
 $\langle\eta_8\rangle$ and $\langle\eta_0\rangle$
 by the factor $m_s/(m_d-m_u)$.
\\

\noindent
(3) Meson condensation breaks CP symmetry (and $U(3)_L\times U(3)_R$ symmetry)
 and induces CP-violating interactions for baryons and mesons.
To study these interactions, we write the baryon chiral Lagrangian at order $O(p^2)$ as
(terms irrelevant in the current discussion are omitted)
\begin{align} 
{\cal L}_{\rm baryons}&={\rm tr}\left[\bar{B}i\gamma^\mu(\partial_\mu B+[\Gamma_\mu,\,B])-M_B \bar{B}B\right]
\nonumber \\
&-\frac{D}{2}{\rm tr}\left[\bar{B}\gamma^\mu\gamma_5\{\xi_\mu,\,B\}\right]
-\frac{F}{2}{\rm tr}\left[\bar{B}\gamma^\mu\gamma_5[\xi_\mu,\,B]\right]
-\frac{\lambda}{2} \ {\rm tr}\left[\xi_\mu\right]{\rm tr}\left[\bar{B}\gamma^\mu\gamma_5B\right]
\nonumber \\
&+b_D \ {\rm tr}\left[\bar{B}\{\chi_+,\,B\}\right]
+b_F \ {\rm tr}\left[\bar{B}[\chi_+,\,B]\right]
+b_0 \ {\rm tr}\left[\chi_+\right]{\rm tr}\left[\bar{B}B\right]+...,
\label{baryon}
\end{align}
 where $B$ represents baryons and $\xi_L,\xi_R$ include mesons as
\begin{align}
B&=\begin{pmatrix} % or pmatrix or bmatrix or Bmatrix or ...
      \frac{1}{\sqrt{2}}\Sigma^0+\frac{1}{\sqrt{6}}\Lambda^0 & \Sigma^+ & p \\
      \Sigma^- & -\frac{1}{\sqrt{2}}\Sigma^0+\frac{1}{\sqrt{6}}\Lambda^0 & n \\
      \Xi^- & \Xi^0 & -\frac{2}{\sqrt{6}}\Lambda^0 \\
   \end{pmatrix},
\\
U&=\xi_R\xi_L^\dagger,
\\
\xi_R&=\xi_L^\dagger.
\end{align}
$\Gamma_\mu$ is a covariant derivative for baryons, $\xi_\mu$ is a combination of meson fields, and $\chi_+$ contains quark masses, which are defined as
\begin{align} % requires amsmath; align* for no eq. number
 \Gamma_\mu&\equiv\frac{1}{2}\xi_R^\dagger(\partial_\mu-i\,r_\mu)\xi_R+\frac{1}{2}\xi_L^\dagger(\partial_\mu-i\,l_\mu)\xi_L,
 \\
\xi_\mu&\equiv i\xi_R^\dagger(\partial_\mu-i\,r_\mu)\xi_R-i\xi_L^\dagger(\partial_\mu-i\,l_\mu)\xi_L,
\\
\chi_+&\equiv2B_0 \ \xi_L^\dagger \ {\rm diag}(m_u,~m_d,~m_s) \ \xi_R+2B_0 \ \xi_R^\dagger \ {\rm diag}(m_u,~m_d,~m_s) \ \xi_L.
\end{align}
$M_B$ is the baryon mass in the chiral limit.
We insert meson VEVs $\langle\pi^0\rangle$, $\langle\eta_8\rangle$, $\langle\eta_0\rangle$ into the baryon chiral Lagrangian Eq.~(\ref{baryon}) and extract CP-violating interaction terms involving neutron $n$. We thus obtain
\begin{align} % requires amsmath; align* for no eq. number
{\cal L}_{\rm baryons}&\supset
\bar{g}_{nn\pi} \ \bar{n}n \,\pi^0_{\rm phys}+\bar{g}_{nn8} \ \bar{n}n\,\eta_{\rm 8phys}
+\bar{g}_{nn0} \ \bar{n}n\,\eta_{\rm 0phys}+\bar{g}_{np\pi}(\bar{p}n\pi^++\bar{n}p\pi^-)
\nonumber\\
&+\bar{g}_{n\Sigma^0 K^0}(\bar{\Sigma}^0n\bar{K}^0+\bar{n}\Sigma^0 K^0)
+\bar{g}_{n\Sigma^- K^+}(\bar{\Sigma}^+nK^-+\bar{n}\Sigma^- K^+)
+\bar{g}_{n\Lambda K}(\bar{\Lambda}n\bar{K}^0+\bar{n}\Lambda K^0),
\label{cpvbaryon}
\end{align}
 where the coupling constants are given by
\begin{align} % requires amsmath; align* for no eq. number
\bar{g}_{nn\pi}&=\frac{B_0}{F_\pi}\left[
4\left\{-b_0(m_u+m_d)-(b_D+b_F)m_d\right\}\frac{\langle\pi^0\rangle}{F_\pi}\right.
\nonumber\\
&\left.+\frac{4}{\sqrt{3}}\left\{b_0(m_d-m_u)+(b_D+b_F)m_d\right\}\left(\frac{\langle\eta_8\rangle}{F_\pi}
+\sqrt{2}\frac{\langle\eta_0\rangle}{F_0}\right)\right],
\\
\bar{g}_{nn8}&=\frac{B_0}{F_\pi}\left[
\frac{4}{\sqrt{3}}\left\{b_0(m_d-m_u)+(b_D+b_F)m_d\right\}\frac{\langle\pi^0\rangle}{F_\pi}\right.
\nonumber\\
&-\frac{4}{3}\left\{b_0(m_u+m_d+4m_s)+b_D(m_d+4m_s)+b_F(m_d-4m_s)\right\}\frac{\langle\eta_8\rangle}{F_\pi}
\nonumber\\
&\left.-\frac{4\sqrt{2}}{3}\left\{b_0(m_u+m_d-2m_s)+b_D(m_d-2m_s)+b_F(m_d+2m_s)\right\}\frac{\langle\eta_0\rangle}{F_0}\right],
\\
\bar{g}_{nn0}&=\frac{B_0}{F_0}\left[
\frac{4\sqrt{2}}{\sqrt{3}}\left\{b_0(m_d-m_u)+(b_D+b_F)m_d\right\}\frac{\langle\pi^0\rangle}{F_\pi}\right.
\nonumber\\
&-\frac{4\sqrt{2}}{3}\left\{b_0(m_u+m_d-2m_s)+b_D(m_d-2m_s)+b_F(m_d+2m_s)\right\}\frac{\langle\eta_8\rangle}{F_\pi}
\nonumber\\
&\left.-\frac{8}{3}\left\{b_0(m_u+m_d+m_s)+b_D(m_d+m_s)+b_F(m_d-m_s)\right\}\frac{\langle\eta_0\rangle}{F_0}\right],
\\
\bar{g}_{np\pi}&=\frac{B_0}{F_\pi}(b_D+b_F)\left[
\sqrt{2}(m_d-m_u)\frac{\langle\pi^0\rangle}{F_\pi}
-\frac{2\sqrt{2}}{\sqrt{3}}(m_u+m_d)\left(\frac{\langle\eta_8\rangle}{F_\pi}+\sqrt{2}\frac{\langle\eta_0\rangle}{F_0}\right)\right],
\\
\bar{g}_{n\Sigma^0K^0}&=\frac{B_0}{F_\pi}(b_D-b_F)\left[
-\frac{1}{2}(3m_d+m_s)\frac{\langle\pi^0\rangle}{F_\pi}
+\frac{1}{2\sqrt{3}}(m_d-5m_s)\frac{\langle\eta_8\rangle}{F_\pi}
+\frac{2\sqrt{2}}{\sqrt{3}}(m_d+m_s)\frac{\langle\eta_0\rangle}{F_0}\right],
\\
\bar{g}_{n\Sigma^-K^+}&=\frac{B_0}{F_\pi}(b_D-b_F)\left[
-\frac{1}{\sqrt{2}}(3m_u+m_s)\frac{\langle\pi^0\rangle}{F_\pi}
+\frac{1}{\sqrt{6}}(-m_u+5m_s)\frac{\langle\eta_8\rangle}{F_\pi}
-\frac{4}{\sqrt{3}}(m_u+m_s)\frac{\langle\eta_0\rangle}{F_0}\right]
\label{nsk}\\
\bar{g}_{n\Lambda K}&=\frac{B_0}{F_\pi}(b_D+3b_F)\left[
-\frac{1}{2\sqrt{3}}(3m_d+m_s)\frac{\langle\pi^0\rangle}{F_\pi}
+\frac{1}{6}(m_d-5m_s)\frac{\langle\eta_8\rangle}{F_\pi}
+\frac{2\sqrt{2}}{3}(m_d+m_s)\frac{\langle\eta_0\rangle}{F_0}\right].
\end{align}
Note in particular that $\langle\pi^0\rangle$ enters into the expression for $\bar{g}_{n\Sigma^-K^+}$ Eq.~(\ref{nsk})
 without the factor of $m_d-m_u$,
 which is allowed because the coupling $\bar{g}_{n\Sigma^-K^+}$ violates isospin.
It follows that $\bar{g}_{n\Sigma^-K^+}$ is enhanced by the factor $m_s/(m_u+m_d)$, as it contains a term 
 $m_s \langle \pi^0\rangle$.

We compare the above meson-VEV-induced CP-violating couplings with those arising from direct hadronic matrix elements
 of ${\cal O}_{1q'q}$ and ${\cal O}_{2q'q}$.
The latter are estimated by na\"ive dimensional analysis~\cite{manohar} as
\footnote{
There are also studies in which the direct hadronic matrix elements are estimated 
 with vacuum saturation approximation~\cite{Khatsimovsky:1987fr,He:1992db,He:1996hb,Hamzaoui:1998yu}
 and with hadron models~\cite{An:2009zh,Xu:2009nt}.
}
 ($B$ and $M$ represent any baryon and meson, respectively)
\begin{align} % requires amsmath; align* for no eq. number
\bar{g}_{BBM}\vert_{\rm direct}&\sim \frac{G_F}{\sqrt{2}}\sum_{i=1,2}\sum_{q,q'}
\vert{\cal C}_{iq'q}\vert\frac{1}{F_\pi}\frac{(4\pi F_\pi)^3}{(4\pi)^2}.
\label{directcontribution}
\end{align}
On the other hand, $\bar{g}_{n\Sigma^-K^+}$ Eq.~(\ref{nsk}), for example, is estimated to be
\begin{align} % requires amsmath; align* for no eq. number
\bar{g}_{n\Sigma^-K^+}\simeq-\frac{B_0}{F_\pi}(b_D-b_F)\frac{m_s}{\sqrt{2}}\frac{\langle\pi^0\rangle}{F_\pi}
&\sim -\frac{G_F}{\sqrt{2}}({\cal C}_{1ud}-{\cal C}_{1du})(b_D-b_F)\frac{4m_s}{\sqrt{2}(m_u+m_d)}\frac{(4\pi F_\pi)^6}{(4\pi)^4F_\pi^3},
\label{mesonvevcontribution}
\end{align}
 where Eq.~(\ref{vevestimate}) and the na\"ive dimensional analysis on $c_3$ Eq.~(\ref{c1c2c3}) are in use.
Noting that $(b_D-b_F)(4\pi F_\pi)\sim1$ holds numerically, we observe that the meson VEV contribution       
 Eq.~(\ref{mesonvevcontribution}) dominates over the direct hadronic matrix element one
 Eq.~(\ref{directcontribution}) by the factor $m_s/(m_u+m_d)$.
This fact allows us to neglect the latter contribution in the rest of the analysis.
\\

\noindent
(4) The neutron EDM receives contributions from baryon-meson loop diagrams involving a CP-violating coupling of Eqs.~(\ref{cpvbaryon}),
 a CP-conserving baryon-meson axial-vector coupling and a photon coupling.
We refer to the loop calculation of Ref.~\cite{guomeissner} performed with infrared regularization~\cite{irr,irrpre},
 from which the neutron EDM, $d_n$, is obtained as
\begin{align} % requires amsmath; align* for no eq. number
d_n\vert_{\rm loop}&=\frac{e}{8\pi^2F_\pi}\left\{
\frac{\bar{g}_{np\pi}}{\sqrt{2}}(D+F)\left(\frac{1}{\bar{\epsilon}}-1-\log\frac{m_\pi^2}{\mu^2}+\frac{\pi m_\pi}{2m_N}\right)\right.
\nonumber\\
&\left.-\frac{\bar{g}_{n\Sigma^-K^+}}{\sqrt{2}}(D-F)\left(\frac{1}{\bar{\epsilon}}-1-\log\frac{m_K^2}{\mu^2}
+\frac{\pi m_K}{2m_N}-\frac{\pi(m_\Sigma-m_N)}{m_K}\right)\right\}.
\label{dnloop}
\end{align}
Here, the divergent part $1/\bar{\epsilon}\equiv1/\epsilon-\gamma_E+\log(4\pi)$ and the scale $\mu$ stem from
 dimensional regularization in $4-2\epsilon$ dimension with mass parameter $\mu$.
In fact, the baryon chiral Lagrangian contains a LEC which cancels the above divergence and whose finite part contributes to the neutron EDM.
The impact of the finite part of the LEC is assessed by na\"ive dimensional analysis~\cite{manohar} as
\begin{align} % requires amsmath; align* for no eq. number
d_n\vert_{\rm LEC}&\sim \frac{G_F}{\sqrt{2}}\sum_{i=1,2}\sum_{q,q'}
\vert{\cal C}_{iq'q}\vert \ e\frac{4\pi F_\pi}{(4\pi)^2}.
\label{directdn}
\end{align}
On the other hand, from Eqs.~(\ref{vevestimate}) and (\ref{nsk}) and the estimate on $c_3$ Eq.~(\ref{c1c2c3}),
 the finite part of the loop contribution Eq.~(\ref{dnloop}) is estimated to be
\begin{align} % requires amsmath; align* for no eq. number
d_n\vert_{\rm loop}&\sim\frac{e}{8\pi^2F_\pi}\frac{\bar{g}_{n\Sigma^-K^+}}{\sqrt{2}}(D-F)
\nonumber\\
&\sim-\frac{e}{8\pi^2F_\pi}\frac{G_F}{\sqrt{2}}({\cal C}_{1ud}-{\cal C}_{1du})(b_D-b_F)\frac{2m_s}{m_u+m_d}\frac{(4\pi F_\pi)^6}{(4\pi)^4F_\pi^3}(D-F).
\label{loopdn}
\end{align}
Since $(b_D-b_F)(4\pi F_\pi)\sim1$ and $D-F\sim1$, we find that the loop contribution Eq.~(\ref{loopdn})
 dominates over the LEC one Eq.~(\ref{directdn}) by the factor $m_s/(m_u+m_d)$.
It is thus justifiable to estimate $d_n$ by simply extracting the finite part of the loop contribution.
We further set $\mu=m_N$, since $m_N$ is a natural cutoff scale, and arrive at
\begin{align} % requires amsmath; align* for no eq. number
d_n&\sim\frac{e}{8\pi^2F_\pi}\left\{
\frac{\bar{g}_{np\pi}}{\sqrt{2}}(D+F)\left(-1-\log\frac{m_\pi^2}{m_N^2}+\frac{\pi m_\pi}{2m_N}\right)\right.
\nonumber\\
&\left.-\frac{\bar{g}_{n\Sigma^-K^+}}{\sqrt{2}}(D-F)\left(-1-\log\frac{m_K^2}{m_N^2}
+\frac{\pi m_K}{2m_N}-\frac{\pi(m_\Sigma-m_N)}{m_K}\right)\right\}.
\label{dn}
\end{align}
\\

Next, we study the case with Peccei-Quinn mechanism.
We incorporate the axion field, $a$, into the meson Lagrangian Eq.~(\ref{meson}) by performing
 $U(3)_A$ chiral rotations to remove the gluon theta term and transform the quark fields as
\begin{align} % requires amsmath; align* for no eq. number
u_L&\to e^{-i\,\alpha_u/2}u_L, \ \ u_R\to e^{i\,\alpha_u/2}u_R, \ \
d_L\to e^{-i\,\alpha_d/2}d_L, \ \ d_R\to e^{i\,\alpha_d/2}d_R,
\nonumber\\
s_L&\to e^{-i\,\alpha_s/2}s_L, \ \ s_R\to e^{i\,\alpha_s/2}s_R,
\label{axion}
\end{align}
 where $\alpha_u,\alpha_d,\alpha_s$ include the axion field $a$ as
\begin{align*} % requires amsmath; align* for no eq. number
\alpha_u&=\frac{m_d m_s}{m_u m_d+m_d m_s+m_s m_u}\left(\frac{a}{f_a}+\bar{\theta}\right),
\\
\alpha_d&=\frac{m_s m_u}{m_u m_d+m_d m_s+m_s m_u}\left(\frac{a}{f_a}+\bar{\theta}\right),
\\
\alpha_s&=\frac{m_u m_d}{m_u m_d+m_d m_s+m_s m_u}\left(\frac{a}{f_a}+\bar{\theta}\right),
\end{align*}
 with $f_a$ denoting the axion decay constant and $\bar{\theta}$ being the genuine theta term.
(With the above choice of $\alpha_u,\alpha_d,\alpha_s$, the axion does not mix with $\pi^0$ or $\eta_8$.)
As a result, the axion field is associated with the quark masses and the coefficients ${\cal C}_{1q'q}$,
 and can thus be implemented in the meson chiral Lagrangian through these terms.
Accordingly, the meson potential Eq.~(\ref{mesonpotential}) is modified to the potential of $\pi^0$, $\eta_8$, $\eta_0$ 
and axion $a$,
\begin{align} % requires amsmath; align* for no eq. number
&V(\pi^0,\eta_8,\eta_0,a)=F_\pi^2B_0\left\{
m_u\cos\left(\frac{\pi^0}{F_\pi}+\frac{\eta_8}{\sqrt{3}F_\pi}+\frac{2\eta_0}{\sqrt{6}F_0}+\alpha_u\right)\right.
\nonumber\\
&+m_d\cos\left(-\frac{\pi^0}{F_\pi}+\frac{\eta_8}{\sqrt{3}F_\pi}+\frac{2\eta_0}{\sqrt{6}F_0}+\alpha_d\right)
\nonumber \\
&\left.+m_s\cos\left(-\frac{2\eta_8}{\sqrt{3}F_\pi}+\frac{2\eta_0}{\sqrt{6}F_0}+\alpha_s\right) \right\}-24\frac{a_0}{F_0^2}(\eta_0)^2
\nonumber\\
&-2c_1\left\{({\cal C}_{1ud}+{\cal C}_{1du})\sin\left(\frac{2\eta_8}{\sqrt{3}F_\pi}+\frac{4\eta_0}{\sqrt{6}F_0}+\alpha_u+\alpha_d\right)\right.
\nonumber\\
&+({\cal C}_{1us}+{\cal C}_{1su})\sin\left(\frac{\pi^0}{F_\pi}-\frac{\eta_8}{\sqrt{3}F_\pi}+\frac{4\eta_0}{\sqrt{6}F_0}+\alpha_u+\alpha_s\right)
\nonumber\\
&\left.+({\cal C}_{1ds}+{\cal C}_{1sd})\sin\left(-\frac{\pi^0}{F_\pi}-\frac{\eta_8}{\sqrt{3}F_\pi}+\frac{4\eta_0}{\sqrt{6}F_0}+\alpha_d+\alpha_s\right)\right\}
\nonumber\\
&-2c_3\left\{({\cal C}_{1ud}-{\cal C}_{1du})\sin\left(-\frac{2\pi^0}{F_\pi}-\alpha_u+\alpha_d\right)+
({\cal C}_{1us}-{\cal C}_{1su})\sin\left(-\frac{\pi^0}{F_\pi}-\frac{\sqrt{3}\eta_8}{F_\pi}-\alpha_u+\alpha_s\right)\right.
\nonumber\\
&\left.+({\cal C}_{1ds}-{\cal C}_{1sd})\sin\left(\frac{\pi^0}{F_\pi}-\frac{\sqrt{3}\eta_8}{F_\pi}-\alpha_d+\alpha_s\right)\right\}.
\label{mesonaxionpotential}
\end{align}
 where it should be reminded that $\alpha_u,\alpha_d,\alpha_s$ are functions of $a$.
The minimization condition for Eq.~(\ref{mesonaxionpotential}) yields meson VEVs $\langle\pi^0\rangle$, $\langle\eta_8\rangle$, $\langle\eta_0\rangle$ and an axion VEV $\langle a\rangle$.
When only the term $({\cal C}_{1ud}-{\cal C}_{1du})$ is non-zero, these VEVs are given by
\begin{align} % requires amsmath; align* for no eq. number
\frac{\langle\pi^0\rangle}{F_\pi}&\simeq
\frac{G_F}{\sqrt{2}}({\cal C}_{1ud}-{\cal C}_{1du})\frac{c_3}{B_0F_\pi^2}
\frac{m_u+m_d+4m_s}{m_um_d+m_dm_s+m_sm_u},
\nonumber\\
\frac{\langle\eta_8\rangle}{F_\pi}&\simeq
\frac{G_F}{\sqrt{2}}({\cal C}_{1ud}-{\cal C}_{1du})\frac{\sqrt{3}c_3}{B_0F_\pi^2}
(m_d-m_u)\frac{1}{m_um_d+m_dm_s+m_sm_u},
\ \ \ \ \
\frac{\langle\eta_0\rangle}{F_0}\simeq0,
\nonumber \\
\frac{\langle a\rangle}{f_a}+\bar{\theta}&\simeq
\frac{G_F}{\sqrt{2}}({\cal C}_{1ud}-{\cal C}_{1du})\frac{2c_3}{B_0F_\pi^2}
(m_d-m_u)\frac{1}{m_um_d}.
\label{vevestimate2}
\end{align}
The VEVs of $\pi^0$ and $\eta_8$ remain of the same order as the case without Peccei-Quinn mechanism,
 and hence they contribute to the neutron EDM in an analogous way.
The axion VEV no longer cancels the genuine $\bar{\theta}$ term and the leftover induces an effective $\bar{\theta}$ term;
 we estimate its contribution by employing the result of Ref.~\cite{edmreview} as
\begin{align} % requires amsmath; align* for no eq. number
d_n\vert_{{\rm ind}\,\bar{\theta}}&=-(2.7\pm1.2)\times 10^{-16}\left(\frac{\langle a\rangle}{f_a}+\bar{\theta}\right) \, e \, {\rm cm}.
\label{dnind}
\end{align}
The final result is the sum of the meson VEV contribution estimated analogously to Eq.~(\ref{dn}), plus Eq.~(\ref{dnind}).

\subsubsection{Other CP-violating operators ${\cal O}_{1q}$, ${\cal O}_{2q}$ and ${\cal O}_3$}
The contributions of the dipole operators in Eq.\ (\ref{Eq:Odipole})
and the Weinberg operator in Eq.\ (\ref{Eq:OWein})
to the neutron EDM can be obtained with
the QCD sum rule.
The former is calculated in Ref.\ \cite{Hisano:2012sc}
while the latter is in Ref.\ \cite{Demir:2002gg},
resulting in the following relations:
\begin{eqnarray}
d_n\vert_{\mathrm{quark}}&=&
0.47d_d-0.12d_u+e(0.18d_d^c-0.18d_u^c-0.008d_s^c),
\label{quarkCont}\\
d_n\vert^{\mathrm{PQ}}_{\mathrm{quark}}&=&
0.47 d_d-0.12d_u+e(0.35d_d^c+0.17d_u^c),
\label{quarkCont2}\\
d_n\vert_{\mathrm{Weinberg}}
&=&\frac{G_F}{\sqrt{2}}eg_s \mathcal{C}_3
\times (10-30)\ \mathrm{MeV},
\label{WeinCont}
\end{eqnarray}
where r.h.s. must be evaluated at $1\ \mathrm{GeV}$.
In Eqs.\ (\ref{quarkCont}, \ref{quarkCont2}), $d_q$ and
$d_q^c (q=u, d, s)$, so-called quark EDM and quark chromo-EDM, are defined as,
\begin{eqnarray}
d_q(\mu)&=&-\frac{G_F}{\sqrt{2}}e\: e_q \mathcal{C}_{1q}(\mu)m_q(\mu),\\
d_q^c(\mu)&=&-\frac{G_F}{\sqrt{2}}\mathcal{C}_{2q}(\mu)m_q(\mu).
\end{eqnarray}
Equations~(\ref{quarkCont}) and (\ref{quarkCont2})
represent the quark EDM contirbutions without and with Peccei-Quinn mechanism, respectively.
For the case without Peccei-Quinn mechanism,
we have taken $\bar{\theta}=0$.
\\

\subsection{Numerical Analysis of Neutron EDM versus $\epsilon'/\epsilon$}

For numerical analysis of $d_n$, we employ the following values:
The chiral-limit pion decay constant $F_\pi$ is obtained from a lattice calculation as $F_\pi=86.8$~MeV~\cite{Fpi}.
$D,F$ have been measured to be $D=0.804$ and $F=0.463$.
For $b_D,b_F$, we quote the result of Ref.~\cite{strangeness,covariantbchpt} with a NLO calculation in Lorentz covariant baryon chiral perturbation theory with decuplet contirbutions, which reads
$b_D=0.161~{\rm GeV}^{-1}$ and $b_F=-0.502~{\rm GeV}^{-1}$.
Since the same calculation formalism, combined with experimental data $\sigma_{\pi N}\simeq 59(7)$~MeV,
 predicts a small value of the strange quark contribution to the nucleon mass $\sigma_s$~\cite{strangeness},
 we infer that these values of $b_D,b_F$ are most robust.
For the quark masses, we adopt lattice results in Ref.~\cite{latticereview}, $m_{ud}(2~{\rm GeV})=3.373$~MeV and $m_s(2~{\rm GeV})=92.0$~MeV,
 and further evaluate QCD five-loop RG evolutions to obtain the masses at 1~GeV in $\overline{MS}$ scheme,
 which are used in our analysis.
Also, we exploit an estimate $m_u/m_d=0.46$~\cite{latticereview}.

The main source of uncertainty in our analysis is the unknown LEC $c_3$ in the meson chiral Lagrangian Eq.~(\ref{meson}).
The other unknown LEC $c_1$ is ineffective, because the Wilson coefficients satisfy $\vert {\cal C}_{1ud}-{\cal C}_{1u}\vert \gg
\vert {\cal C}_{1ud}+{\cal C}_{1u}\vert,\,\vert{\cal C}_{1sq}\vert,\,\vert{\cal C}_{qs}\vert$.
Our calculations of loop-induced $d_n$ Eq.~(\ref{dn}) and axion-induced $d_n$ Eq.~(\ref{dn}) are hence proportional to $c_3$
 and subject to $O(1)$ uncertainty originating from its na\"ive dimensional analysis Eq.~(\ref{c1c2c3}).
The fact that our results depend only on one LEC $c_3$ is good news, because it excludes the possibility of accidental cancellation between contributions with different LECs.
Another source of uncertainty is the renormalization scale $\mu$ in the loop calculation Eq.~(\ref{dnloop}),
 but this is subdominant compared to the uncertainty of $c_3$.
\\

In the analysis, the ratio of the bifundamental scalar VEVs is again fixed as $\tan\beta=m_b/m_t$.
The values of the new CP phases $\phi_c, \phi_t, \psi_d, \psi_s, \psi_b, \alpha$ are randomly generated.
We find that the contribution of the Weinberg operator is suppressed by roughly $10^{-7}-10^{-9}$ compared with that of the four-quark operators, and thus we neglect it in the analysis.

\begin{figure}[H]
\centering
\includegraphics[width=12.0cm]{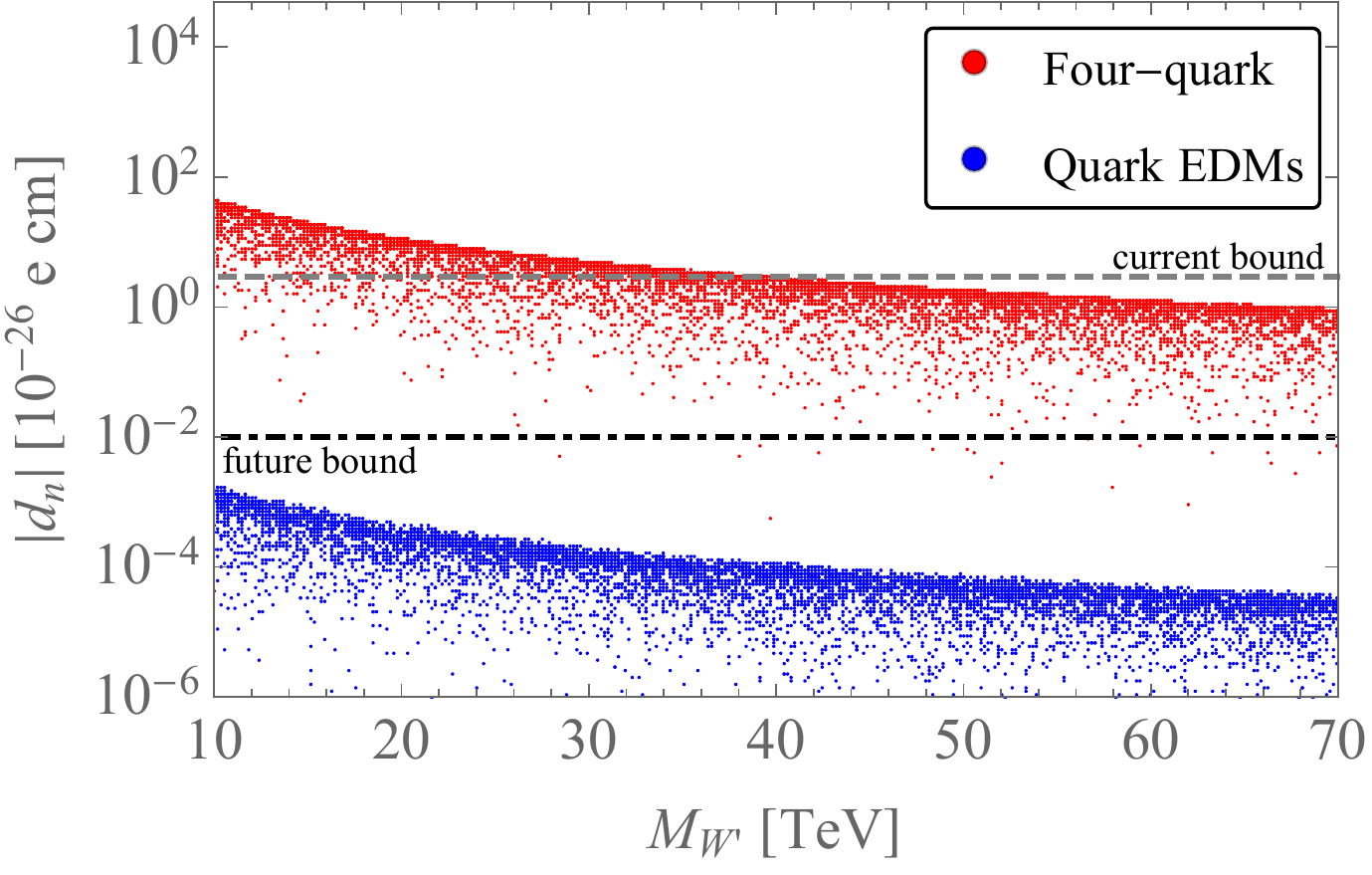}
  \caption{Prediction for the neutron EDM
  in the case with $\bar{\theta}=0$ without Peccei-Quinn mechanism.
  Only the contributions of four-quark operators and
  quark-level EDMs including both quark EDM and chormo-EDM are shown.
  A dashed line represents the current bound on the neutron EDM \cite{nedm}, while
  a dashed dotted line stands for the future bound \cite{Kumar:2013qya}.}
\label{Fig:3}
\end{figure}

\begin{figure}[H]
\centering
\includegraphics[width=12.0cm]{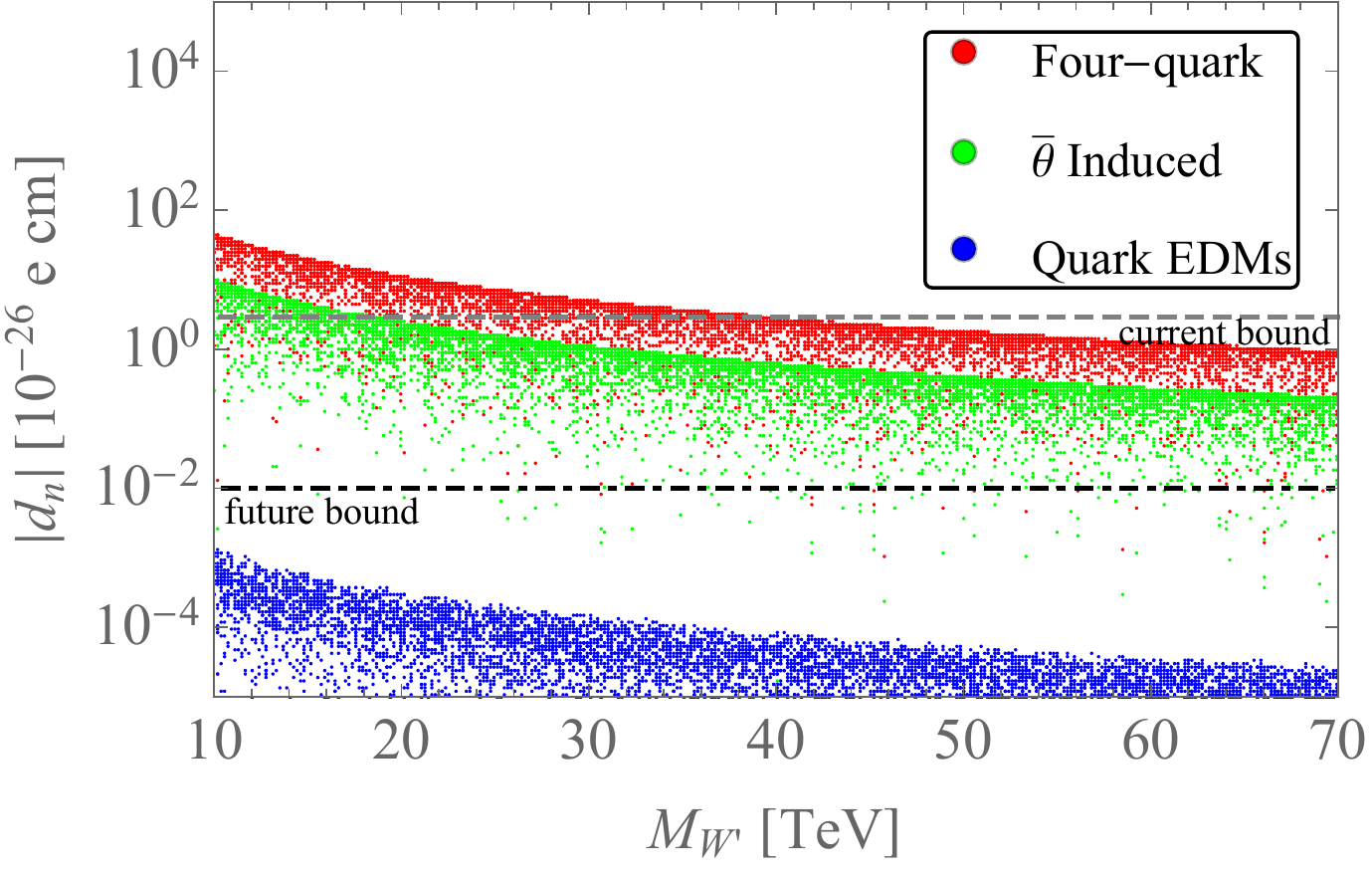}
  \caption{Comparison between the contribution of the induced $\bar{\theta}$
  to the neutron EDM and others in the presence of
  Peccei-Quinn mechanism.
  A gray dashed line represents the current bound of EDM \cite{nedm} while
  a black dashed dotted line stands for the future bound \cite{Kumar:2013qya}.}
\label{Fig:4}
\end{figure}

First, we show the numerical result for the neutron EDM
without the constraint from $\epsilon^\prime/\epsilon$ in Fig.\ (\ref{Fig:3}).
One observes that the contribution of the four-quark operators is dominant over
that of the quark EDMs.
As stated previously, an effective $\bar{\theta}$ term
is induced in the presence of Peccei-Quinn mechanism.
In Fig.\ \ref{Fig:4},
we additionally show the numerical prediction based on Eq.\ (\ref{dnind}).
One finds that the induced $\bar{\theta}$ gives subleading contribution to the neutron EDM.
\par
\begin{figure}[H]
\centering
\includegraphics[width=13.5cm]{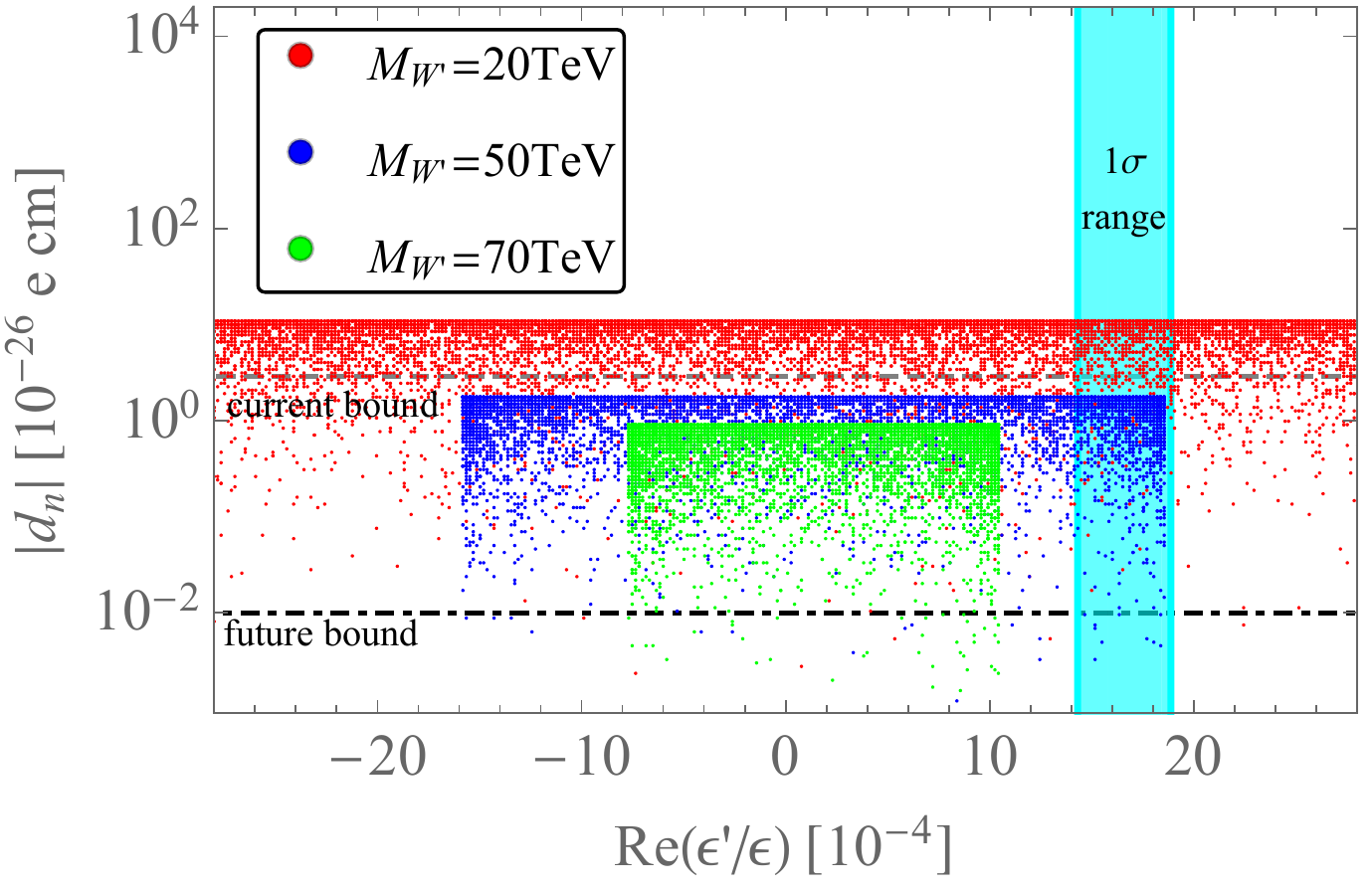}
  \caption{Correlation plot for the direct CP violation in $K\to\pi\pi$ decay and the neutron EDM 
  in the case with $\bar{\theta}=0$ without
  Peccei-Quinn mechanism.
  A gray dashed line and a black dashed dotted line
  represent the current \cite{nedm} and the future \cite{Kumar:2013qya}
  bounds on the neutron EDM,
  while a cyan band stands for the $1\sigma$ range of the direct CP violation in $K\to\pi\pi$ decay
  obtained from PDG \cite{Patrignani:2016xqp}.}
\label{Fig:5}
\end{figure}
\begin{figure}[H]
\centering
\includegraphics[width=13.5cm]{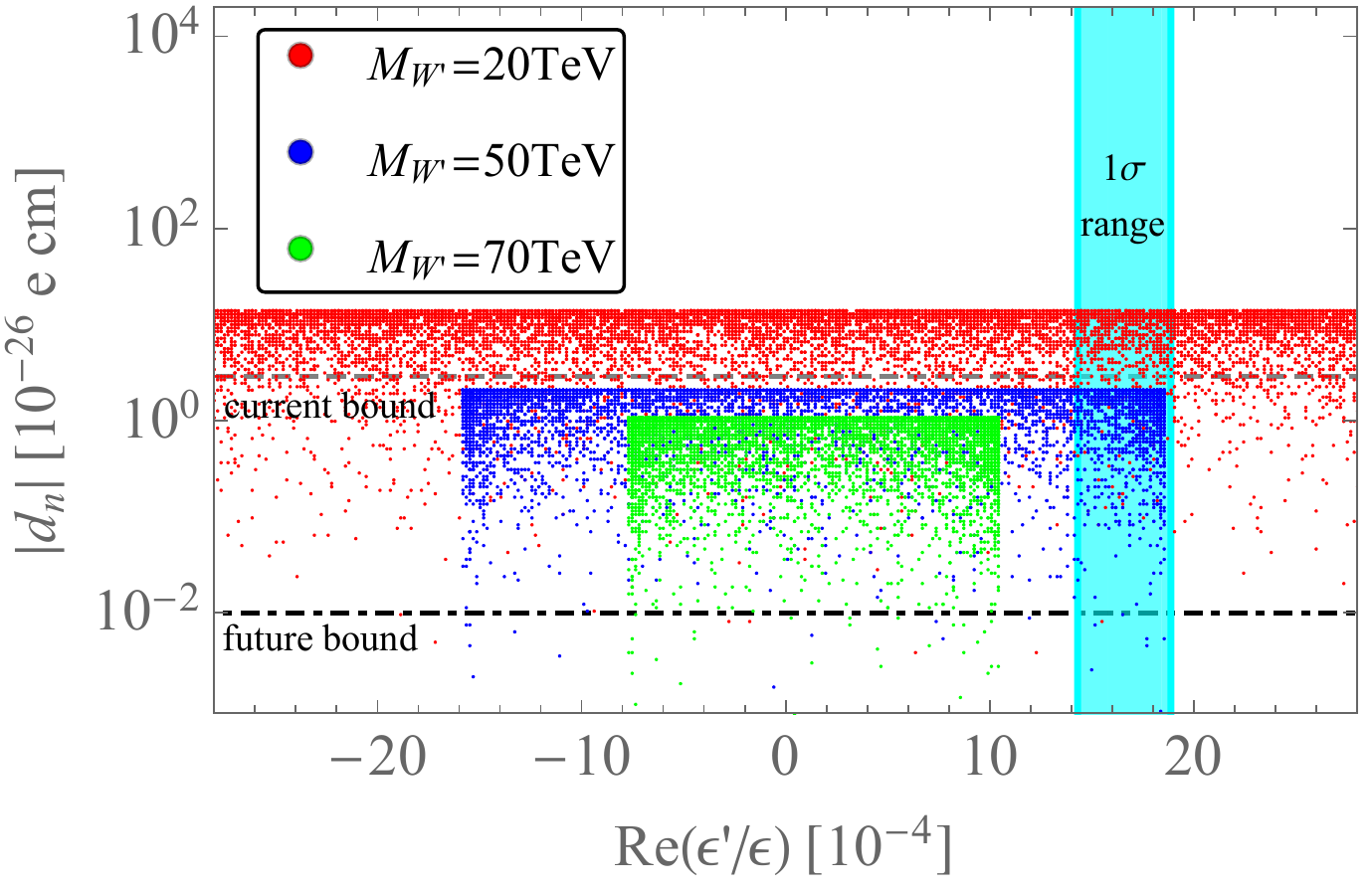}
  \caption{The same figure as Fig.\ \ref{Fig:5}
  with Peccei-Quinn mechanism.}
\label{Fig:6}
\end{figure}

Next, the correlated prediction for $|d_n|$ and $\mathrm{Re}(\epsilon^\prime/\epsilon)$ is presented in Figs.\ \ref{Fig:5} and \ref{Fig:6} in the cases without and with
Peccei-Quinn mechanism, respectively.
Here, small contributions from the quark EDMs are neglected.
The cases with and without Peccei-Quinn mechanism yield almost identical results
 because the induced $\bar{\theta}$ has a subdominant effect, as seen in Fig.\ \ref{Fig:4}.
We observe that $M_{W^\prime}=20\ \mathrm{TeV}$ and $50\ \mathrm{TeV}$
can be consistent with the data on $\mathrm{Re}(\epsilon^\prime/\epsilon)$ at $1\sigma$ level,
whereas the case with $M_{W^\prime}=70\ \mathrm{TeV}$ cannot explain it.
However, the case with $M_{W^\prime}=20\ \mathrm{TeV}$ has already been excluded by the current bound on the neutron EDM,
 and only $M_{W^\prime}=50\ \mathrm{TeV}$ can be compatible with the neutron EDM bound and the data on
 $\mathrm{Re}(\epsilon^\prime/\epsilon)$.
Figs.\ \ref{Fig:5} and \ref{Fig:6} further inform us that
 almost all parameter points that account for the $\mathrm{Re}(\epsilon^\prime/\epsilon)$ data
 will be covered by future neutron EDM searches \cite{Kumar:2013qya}.
Therefore, unless the tree-level $\bar{\theta}$ in the case without Peccei-Quinn mechanism miraculously cancels the contribution of the model,
 we anticipate the discovery of the neutron EDM in the near future.

\section{Summary and Discussions}
\label{Summary}

We have addressed the $\epsilon'/\epsilon$ anomaly 
 in the $SU(2)_L\times SU(2)_R\times U(1)_{B-L}$ gauge extension 
 of the SM with charge symmetry.
Since the charge symmetry gives strong restrictions on the mixing matrix for right-handed quarks,
 $\epsilon'/\epsilon$ can be evaluated only in terms of two new CP phases $\alpha-\psi_d$ and $\alpha-\psi_s$,
 the mass of $W'$ gauge boson (mostly composed of $W_R$), and the bifundamental scalar VEV ratio $\tan\beta$.
By fixing $\tan\beta$ at its natural value $m_b/m_t$, and by randomly varying $\alpha-\psi_d$ and $\alpha-\psi_s$,
 we have shown that $M_{W'}<58$~TeV must be satisfied to account for the experimental value of $\epsilon'/\epsilon$
 at 1 $\sigma$ level.

Next, we have made a prediction for the neutron EDM $d_n$ 
 when the $SU(2)_L\times SU(2)_R\times U(1)_{B-L}$ model with charge symmetry solves the $\epsilon'/\epsilon$ anomaly.
We have investigated the contribution of meson condensates induced by four-quark operators,
 and revealed that the $\pi^0$ VEV dominantly contributes to the neutron EDM, whose impact is 
 enhanced by $m_s/(m_u+m_d)$ compared to other contributions.
This enhancement is attributable to the isospin violating coupling of $W'$ gauge boson, which allows the $\pi^0$ VEV
 to arise without the factor of $m_d-m_u$.
Additionally, we have found that the induced $\bar{\theta}$ term in the presence of Peccei-Quinn mechanism yields only a subleading effect on $d_n$.
On the basis of the above observations, we have shown that
 the $\epsilon'/\epsilon$ anomaly can be explained without conflicting the current experimental bound on $d_n$,
 and that the parameter space where the $\epsilon'/\epsilon$ data are accounted for
 will be almost entirely covered by future experiments~\cite{Kumar:2013qya}.

We comment on the constraint from Re($\epsilon$) on the model.
Since $W'$ gauge boson contributes to $\Delta F=2$ processes only at loop levels,
 for $M_{W'}>20$~TeV, its contribution to Re($\epsilon$) is safely below the experimental bound~\cite{Bertolini:2014sua}.
However, the heavy neutral scalar particles coming from the bifundamental scalar induce $\Delta F=2$ processes at tree level.
Since their mass is of the same order as or below $M_{W'}$ if there is no fine-tuning in the scalar potential,
 these particles may lead to a tension with the data on Re($\epsilon$)~\cite{Bertolini:2014sua}
 (constraint from Re($\epsilon$) on general left-right models is found in Ref.~\cite{Blanke:2011ry},
 and that on the model with left-right parity is in Ref.~\cite{Haba:2017jgf})
 (for early studies on the Re($\epsilon$) constraint, see, e.g., Ref.~\cite{Kiers:2002cz}).

\section*{Acknowledgement}

The authors would like to thank Monika Blanke, Andrzej Buras, Antonio Pich and Amarjit Soni for valuable comments.
This work is partially supported by Scientific Grants by the Ministry of Education, Culture, Sports, Science and Technology of Japan (Nos. 24540272, 26247038, 15H01037, 16H00871, and 16H02189).
\\

\section*{Appendix A}
\begin{align} % requires amsmath; align* for no eq. number
O_1&=(\bar{s}_\alpha u_\beta)_L(\bar{u}_\beta d_\alpha)_L, \ \ \ 
O_2=(\bar{s}u)_L(\bar{u}d)_L,
\\
O_{1c}&=(\bar{s}_\alpha c_\beta)_L(\bar{c}_\beta d_\alpha)_L, \ \ \ 
O_{2c}=(\bar{s}c)_L(\bar{c}d)_L,
\\
O_1^{RL}&=(\bar{s}_\alpha u_\beta)_R(\bar{u}_\beta d_\alpha)_L, \ \ \ 
O_2^{RL}=(\bar{s}u)_R(\bar{u}d)_L,
\\
O_{1c}^{RL}&=(\bar{s}_\alpha c_\beta)_R(\bar{c}_\beta d_\alpha)_L, \ \ \ 
O_{2c}^{RL}=(\bar{s}c)_R(\bar{c}d)_L,
\\
O_3&=\sum_{q=u,d,s}(\bar{s}d)_L(\bar{q}q)_L, \ \ \
O_4=\sum_{q=u,d,s}(\bar{s}_\alpha d_\beta)_L(\bar{q}_\beta q_\alpha)_L,
\\
O_5&=\sum_{q=u,d,s}(\bar{s}d)_L(\bar{q}q)_R, \ \ \
O_6=\sum_{q=u,d,s}(\bar{s}_\alpha d_\beta)_L(\bar{q}_\beta q_\alpha)_R,
\\
O_7&=\frac{3}{2}\sum_{q=u,d,s}(\bar{s}d)_L e_q(\bar{q}q)_R, \ \ \
O_8=\frac{3}{2}\sum_{q=u,d,s}(\bar{s}_\alpha d_\beta)_L e_q(\bar{q}_\beta q_\alpha)_R,
\\
O_9&=\frac{3}{2}\sum_{q=u,d,s}(\bar{s}d)_L e_q(\bar{q}q)_L, \ \ \
O_{10}=\frac{3}{2}\sum_{q=u,d,s}(\bar{s}_\alpha d_\beta)_L e_q(\bar{q}_\beta q_\alpha)_L,
\\
O_g&=\frac{g_s}{8\pi^2}m_s \bar{s}\sigma_{\mu\nu}G^{a\mu\nu}T^aP_L d,
\ \ \
O_\gamma=\frac{e}{8\pi^2}m_s \bar{s}\sigma_{\mu\nu}F^{\mu\nu}P_Ld,
\end{align}
 where $(\bar{q} q')_L\equiv\bar{q}\gamma_\mu(1-\gamma_5)q'$ and $(\bar{q} q')_R\equiv\bar{q}\gamma_\mu(1+\gamma_5)q'$,  $\alpha,\beta$ are color indices, and color summation is taken in each quark bilinear unless $\alpha,\beta$ are displayed.
$e_u=2/3$ and $e_d=e_s=-1/3$.
The operators $O'_i$, $O_j^{LR}$ are obtained by interchanging $L\leftrightarrow R$ in the corresponding operators.
\\

\section*{Appendix B}

The loop functions in the main text are defined as follows:
\begin{align*} % requires amsmath; align* for no eq. number
F_1(x)&=\frac{x(-18+11x+x^2)}{12(1-x)^3}+\frac{x^2(-15+16x-4x^2)}{6(1-x)^4}\log x+\frac{2}{3}\log x+\frac{2}{3},
\\
F_2(x)&=\frac{x(2+5x-x^2)}{4(1-x)^3}+\frac{3x^2\log x}{2(1-x)^4},
\\
F_3(x)&=\frac{4+x+x^2}{2(1-x)^2}+\frac{3x\log x}{(1-x)^3},
\\
E_{1d}(x)&=\frac{25x^2-19x^3}{36(1-x)^3}+\frac{x^2(6+2x-5x^2)}{18(1-x)^4}\log x+\frac{4}{9}\log x+\frac{4}{9},
\\
E_{2d}(x)&=\frac{x(7-5x-8x^2)}{12(1-x)^3}+\frac{x^2(2-3x)}{2(1-x)^4}\log x,
\\
E_{3d}(x)&=\frac{20-31x+5x^2}{6(1-x)^2}+\frac{x(2-3x)}{(1-x)^3}\log x,
\\
E_{3u}(x)&=\frac{8-16x+2x^2}{3(1-x)^2}+\frac{x(1-3x)}{(1-x)^3}\log x.
\end{align*}
\\

\section*{Appendix C}

\begin{align} % requires amsmath; align* for no eq. number
{\cal O}_{1q}&=-\frac{e}{2}e_q\,m_q\bar{q}\sigma_{\mu\nu}i\gamma_5q\,F^{\mu\nu},
\ \ \
{\cal O}_{2q}=-\frac{g_s}{2}\,m_q\bar{q}\sigma_{\mu\nu}i\gamma_5T^aq\,G^{a\,\mu\nu},\label{Eq:Odipole}
\\
{\cal O}_3&=-\frac{1}{6}f^{abc}\epsilon^{\mu\nu\rho\sigma}\eta^{\tau\upsilon}
G^a_{\tau\mu}G^b_{\nu\rho}G^a_{\sigma\upsilon},\label{Eq:OWein}
\\
{\cal O}_{4q}&=\bar{q}q \ \bar{q}i\gamma_5q, \ \ \ {\cal O}_{5q}=\bar{q}\sigma_{\mu\nu}q \ \bar{q}\sigma^{\mu\nu}i\gamma_5q,
\\
{\cal O}_{1q'q}&=\bar{q}'q' \ \bar{q}i\gamma_5q,
\ \ \ {\cal O}_{2q'q}=\bar{q}'_\alpha q'_\beta \ \bar{q}_\beta i\gamma_5q_\alpha,
\\
{\cal O}_{3q'q}&=\bar{q}'\sigma^{\mu\nu}q' \ \bar{q}\sigma_{\mu\nu}i\gamma_5q,
\ \ \ {\cal O}_{4q'q}=\bar{q}'_\alpha \sigma^{\mu\nu}q'_\beta \ \bar{q}_\beta\sigma_{\mu\nu} i\gamma_5q_\alpha,
\end{align}
 where $q',q=u,d,s$ and $q'\neq q$.
 $\alpha,\beta$ are color indices, and color summation is taken in each quark bilinear unless $\alpha,\beta$ are displayed.

\end{document}